\documentclass[USenglish, twocolumn]{article}

\ifx\directlua\undefined\ifx\XeTeXcharclass\undefined
  \usepackage[utf8]{inputenc}                           
  \else\RequirePackage[no-math]{fontspec}[2017/03/31]\fi 
  \else\RequirePackage[no-math]{fontspec}[2017/03/31]\fi 
\usepackage[sort&compress,square,numbers]{natbib}
\usepackage{graphicx}
\usepackage{amsmath}
\usepackage[dvipsnames]{xcolor}
\usepackage{xr}
\title{
Toward Reliable Spectroscopic Analysis of Reaction Kinetics in Polaritonic Chemistry}

\author{Robrecht M.~A.~Vergauwe$^{1*}$, J.~Jussi Toppari$^{2*}$, Gerrit Groenhof$^{1*}$ \\
\\
$^1${\em Nanoscience Center and Department of Chemistry, University of Jyv\"{a}skyl\"{a}, }\\ {\em P.O. Box 35, 40014 Jyv\"{a}skyl\"{a}, Finland} \\
$^2${\em Nanoscience Center and Department of Physics, University of  Jyv\"{a}skyl\"{a}, }\\ {\em P.O. Box 35, 40014 Jyv\"{a}skyl\"{a}, Finland }\\
\\
robrecht.vergauwe@gmail.com\\
j.jussi.toppari@jyu.fi; 0000-0002-1698-5591 \\
gerrit.x.groenhof@jyu.fi; 0000-0001-8148-5334}

\begin{document}

\maketitle

\abstract{
Recent reports suggest that chemical reaction rates can change when reactants are placed inside an optical cavity. These effects have been attributed to the hybridization of molecular vibrational modes with cavity modes into polaritons, but the underlying mechanism remains debated. Recently, attempts to reproduce the key experiments have sometimes failed, which poses also ambiguity and impedes the determination of the possible mechanism.

Without a reliable theoretical framework, polaritonic chemistry -- which seeks to use optical resonators as catalysts to control reactions -- has reached a pivotal stage. Standardized protocols for reproducible cavity experiments are therefore urgently needed. Here, we identify pitfalls in approaches that monitor reaction progress with UV/Vis spectroscopy. Using the Transfer Matrix Method, we analyze a model pseudo-first-order reaction and assess how transient cavity thickness variations, cavity inhomogeneity, and fitting protocols influence the extracted rate constant. We find that changes in cavity thickness upon reactant introduction can strongly distort apparent kinetics when monitoring at a single wavelength, an artifact that can be mitigated by spectral smoothing. Additionally, we demonstrate that, unlike in many previous studies, the asymptotic extinction should be treated as a fitting parameter rather than fixed to the final experimental value. By identifying these pitfalls, our work lays the foundation for more robust analyses and reliable measurements in polaritonic chemistry.
}


\section{Introduction} 

It is possible to create hybrid states of light and matter, known as polaritons, using an IR-active vibration of an organic substance and a photonic cavity -- a structure that confines electromagnetic fields of a certain frequency in a limited region in space~\cite{KeiReview, SimpkinsReview, XiongReview}. If 1) the cavity frequency matches the energy of the vibration, i.e., resonance occurs, and 2) the molecule-cavity EM fields interaction, i.e., photon absorption/re-emission, is stronger than any other loss mechanism such as photon leakage or dissipation by the molecule, the so-called strong coupling regime is reached, and the wavefunctions of molecule and cavity field hybridize, forming the polaritons~\cite{BorjessonReview, BarnesTormaReview}. This phenomenon is called vibrational strong coupling (VSC). Because this entails a direct modification of the molecular wavefunction, the question arises if such hybridization could affect the physico-chemical properties of the system. 

To address this question, several groups have begun measuring these properties~\cite{KeiReview, SimpkinsReview, XiongReview}. These endeavors have produced fascinating claims, such as reaction rate acceleration~\cite{Hiura2018, JinoGeorge1, JinoGeorge2, Jaime} and decelerations~\cite{EbbesenScience, Hirai, Vergauwe, Nagarajanpud, SimpkinsScience, Moran2025}, chemical equilibrium shifts \cite{Yantao, EbbesenNMR, Jayachandran2025, KripaPorphyrins, Kripa, Kripa2025} and the transformation of an insulator into a conductor~\cite{AnoopConductivity}, raising both excitement and controversy~\cite{Thomas2024,WeiXiong,Imperatore2021}. Yet the central mystery remains unresolved: how, and to what extent, does VSC actually control chemical reactions? Existing theoretical models explain fragments of the observations~\cite{FeistSN2, KowalewskiTheory, SchaferTheory, VendrellTheory, HuoTheory} but fail to provide predictive power~\cite{YuenZhouReview, HsuReview, VidalFeistReview}.

On the experimental side, the role of photonic cavities and polaritons in catalysis remains highly debated. George and co-workers initially reported a tenfold acceleration of the fluoride-mediated cleavage of para-nitrophenol acetate (PNPA) under vibrational strong coupling (VSC) \cite{JinoGeorge1}, and a 10\,000\,-fold acceleration of cyanide hydrolysis \cite{Hiura2018}. However, these results could not be reproduced by others~\cite{WeiXiong,Imperatore2021,JinoGeorge2}, nor by the same authors, who later observed only a 2.5\,–\,3\,-fold rate increase — much smaller and within the typical uncertainties of kinetic measurements~\cite{JinoGeorge2}. Gomez Rivas and colleagues likewise found only a very modest rate changes in a similar cavity setup, though they reported a 2.7-fold enhancement when the reaction was coupled to plasmonic nanoparticle arrays \cite{Jaime}. These discrepancies place polaritonic chemistry at a pivotal stage, underscoring the need for careful experiments to resolve the current controversies.

Most of the polaritonic chemistry experiments have been performed in microfluidic Fabry-P\'{e}rot (FP) micro-cavities, built from commercial IR transmission cells with $\mu$m optical path lengths~\cite{NagarajanReview,KeiReview, EbbesenScience, SimpkinsScience}. An ideal Fabry-P\'erot structure consists of two planar mirrors facing each other in a perfectly parallel manner and supports a series of regularly spaced transmission modes. In practice, microfluidic VSC cavities have thin $\sim$10 nm Au mirrors and display clear transmission modes in the infrared. In the UV/Vis, these cavities have only an oscillatory transmission spectrum of higher order cavity modes enveloped by the imprint of the intraband transitions of Au~\cite{Vergauwe, JinoGeorge2, WeiXiong}. Recently, Michon and Simpkins demonstrated that optical measurements of organic dyes in such non-ideal Fabry-P\'{e}rot cavities can give rise to artifacts because of the higher order mode fringes in the UV/Vis transmission spectra.~\cite{Simpkins}. A dye with a peak at 600 nm will display a shifted peak position when there is sufficient broadening due to cavity width inhomogeneity, i.e. the cavity has a range of different cavity widths in the area being probed, for example due to misalignment or bending of the mirrors under pressure~\cite{Simpkins}. Such width inhomogeneity can have important implications for reaction kinetic measurements on (pro-)chromophores like PNPA, and one cannot assume anymore that chemical kinetics in a Fabry-P\'erot structure can be directly extracted from UV/Vis data, in particular if only a single wavelength is analyzed. 

In addition to cavity width inhomogeneity, characterized by a static distribution of cavity widths, we identified another issue during our VSC chemistry experiments: microfluidic VSC cavities display almost always some degree of contraction after injection, i.e., the cavity width does not remain fixed but instead decreases during the course of the reaction measurement. The impact of this contraction on reaction rate estimation has not yet been investigated. Given that a residual cavity mode structure is visible in cavity UV/Vis spectra, failing to account for spectral effects of transient cavity contraction can also impact the reaction rate estimation. The unavoidable presence of cavity width distributions and temporal cavity width changes — static and dynamic phenomena, respectively — forces us to consider the extent to which reaction rates can be reliably estimated from Fabry–P\'erot UV/Vis data. 

To address these issues, we introduce a controllable model system for systematically investigating how various optical artifacts influence a (pseudo-)first-order reaction of a PNPA-type pro-chromophore using transfer matrix method (TMM) simulations. Beyond static cavity-width inhomogeneity, the model includes dynamic cavity contraction — motivated by experimental observations — as well as different procedures for processing raw absorption spectra and fitting reaction rates, including the determination of the asymptotic extinction, $\text{Ext}_\infty$, which is shown to be of critical importance. We focus on (pseudo-)first-order kinetics, as this regime predominates in studies of bimolecular reactions ($A + B \rightarrow C$) in physical and polaritonic chemistry. Since our model excludes any coupling between molecular vibrations and the cavity field, any apparent changes in reaction rate inferred from the simulated spectroscopic signals must arise solely from optical artifacts or analysis procedures. Our simulations reveal that cavity thickness fluctuations upon reactant introduction can significantly distort kinetic traces when monitoring reaction progress at a \emph{single} wavelength. Although spectral smoothing and integration can mitigate this effect, care must be taken in interpreting the resulting kinetics. We also show that the asymptotic extinction should be treated as a fitting parameter rather than fixed to the final experimental value. Finally, we discuss how the artifacts identified here may have influenced prior analyses of PNPA hydrolysis under VSC conditions.

\section{Materials and Methods}

Fabry-P\'erot structures are simulated using the standard Transfer Matrix Method (TMM) on the following stack structure: air ($n=1$) / Au mirror / medium layer / Au mirror / air ($n=1$)
~\cite{FundPhot}. The two Au mirrors layers are modeled as perfectly planar 10 nm thick Au layers with a complex refractive index taken from Rosenblatt \textit{et al.}~\cite{Rosenblatt}. The medium layer thickness is 5 $\mu$m, except when simulating the experiments of Lather \textit{et al.}, Singh \textit{et al.}~and Weisehan \textit{et al.}, in which the width was 18-22~$\mu$m~\cite{WeiXiong, JinoGeorge1, JinoGeorge2} (see SI). To model a reaction that produces a \textit{para}-nitrophenolate-type chromophore within the organic medium layer, the time-dependent dielectric function of this layer was constructed from a fully Kramers-Kronig compatible Gaussian broadened Lorentz oscillator model, $\epsilon_{\text{absorber}}(\omega)$~\cite{Orosco}, parameterized for an absorption peak at 400~nm with a linewidth of 1000~$\text{cm}^{-1}$ at full-width at half maximum (FWHM, see Supplementary Information (SI) for more details), and a background refractive index $n_r$ of 1.3, similar to methanol and ethanol. The amplitude factor of this model, $\tilde{a} = 1.2\times10^{-4}$, was multiplied with a coefficient $0 \leq c_\text{r}(t) \leq 1$ that reflects the progression of the (pseudo-)first order reaction with rate $k_\text{reaction}$:
\begin{equation}
\epsilon_\text{medium}(\omega,t) = \epsilon_0 + c_r(t)\cdot\tilde{a}\cdot\epsilon_{\text{absorber}}(\omega) \label{eq:progression}
\end{equation}
with 
\begin{equation}
c_r(t) = 1 - e^{ - k_\textrm{reaction} t}\label{eq:concentration}.
\end{equation}
Thus, the time-dependent concentration of the reaction product $C(t)$ is proportional to $c_r(t)\cdot\tilde{a}$.

Unless specified otherwise, we use reaction rate of $k_\mathrm{reaction} = 2\times 10^{-4}~\text{s}^{-1}$ and the simulations were carried out for 8000~s with time steps of 10~s. The time-dependent  transmission spectra $T_\mathrm{sim}(\lambda,t)$, defined as the fraction of power transmitted, were computed for the wavelengths $\lambda$ from 390 to 850~nm with a sampling resolution of 1~nm (see SI for comment on light polarization).
The time-dependent cavity extinction spectra for unpolarized light in units of optical density (OD) were subsequently computed as:
\begin{equation}
\text{Ext}(\lambda,t) = -\log_{10}\left( T_\text{sim}(\lambda,t) \right)\label{eq:cavityext}
\end{equation}
This way, we mimic previous VSC experiments, in which the reaction progress was tracked using commercial UV/Vis spectrophotometers with unpolarized light sources.

To calculate the reference time-dependent UV/Vis extinction spectrum of a reaction taking place in a standard transmission measurement cell, a TMM calculation is performed on a stack lacking the Au mirrors: air ($n=1$) / medium layer / air ($n=1$). The reaction and the calculation of the cell extinction spectra were treated identically to the procedure described above for the cavity.

To account for cavity width inhomogeneity in our simulations, we followed the approach of Michon and Simpkins~\cite{Simpkins} and assumed that the probe beam samples a Gaussian distribution of cavity widths, with a standard deviation given by $\sigma_\textrm{FWHM} = L_\textrm{FWHM} / \left(2\sqrt{2\ln{2}}\right)$, where $L_\textrm{FWHM}$ denotes the full width at half maximum of the distribution. The widths, $L_i$, were then deterministically sampled from this distribution, and the transmission spectrum of the Fabry–P\'{e}ot structure was computed as a weighted sum of the spectra corresponding to the individual cavity widths (see SI):
\[
T = \frac{1}{N}\sum_i^N w(L_i)\times T(L_i)
\]
with $T(L_i)$ being the transmission of a cavity with width $L_i$, and weight
\[
w(L_i) = \frac{1}{\sqrt{2\pi} \sigma _\textrm{FWHM}}\cdot\exp{ \left[ \frac{- ( L-L_i )^2 }{ 2\sigma _\textrm{FWHM}^2} \right] }
\]
We computed $N=17$ transmission spectra, with cavity lengths evenly distributed between $L-4L_\textrm{FWHM}$ and $L+4L_\textrm{FWHM}$ around the average cavity length $L$. In SI we demonstrate the much higher performance, i.e. reduced noise and computation cost, of deterministic sampling of the cavity width distribution compared to random sampling (Fig.~S12).

We tested four different techniques for extracting the reaction progress from the calculated extinction spectra (for details see SI).
\begin{enumerate}
    \item \textbf{Savitzky-Golay filtering}~\cite{SciPy-NMeth} on a 51 data point domain.
    \item \textbf{Rolling-average filtering} on a domain of 41 points.
    \item \textbf{Integration} of the spectra over spectral regions between 405-415, 395-425 or 400-450 nm. The integrals were divided by the length of the spectral region to facilitate direct comparison.
    \item \textbf{Integration after rolling-average filtering}. Integration over the spectral region between 410 and 430 nm after application of a rolling-average filter (2).
\end{enumerate}

Progress of the reaction was either inferred from the differential extinction at 410~nm with respect to the extinction at time 0, i.e. $\Delta \text{Ext}(410\text{~nm},t)=\text{Ext}(410\text{~nm},t)-\text{Ext}(410\text{~nm},0\text{~s})$ (equation~\ref{eq:cavityext}) or from the integrated differential extinction spectrum. All the filtering/integration/averaging were done to the original extinction spectra, i.e., before the subtraction of $\text{Ext}(410\text{~nm},t=0\text{~s})$.

Direct fitting of the reaction progress to the differential-extinction time traces was performed using one of two exponential models:
\begin{equation} \label{eq:ExpSimple}
\Delta \text{Ext}(t) = \text{Ext}_\infty \left( 1 - e^{-k_\mathrm{fit} t} \right),
\end{equation}
or
\begin{eqnarray} \label{eq:ExpFull}
\Delta \text{Ext}(t) \!\!&=&\!\! \text{Ext}_0 \left[ 1 - e^{-k_\mathrm{fit}(t - t_0)} \right] + \text{Ext}_\mathrm{bck} \\
\text{Ext}_\infty \!\!&=&\!\! \text{Ext}_0 + \text{Ext}_\mathrm{bck}. \nonumber
\end{eqnarray}
Here, $k_\text{fit}$ is the fitted rate constant; any deviation from the expected reaction rate $k_\text{reaction}$ therefore indicates an artifact rather than a physical change in kinetics. Equation \ref{eq:ExpSimple} is used for simulations of an ideal, homogeneous cavity (Figures 1 and S1), where neither time delays nor background extinction are present. When cavity contraction is included (Figures 3, S5, and S6), the more general four-parameter form (Equation \ref{eq:ExpFull}) is employed. This expression allows the exponential to be translated in both time and extinction via the parameters $t_0$ and $\text{Ext}_\text{bck}$ and corresponds to the model commonly used to analyze experimental data, where finite response times and background signals are unavoidable. Importantly, our simulations do not contain such effects -- the reaction begins at $t=0$ with zero initial product concentration. We nevertheless use the full model to evaluate how spectral changes induced by cavity contraction during the measurement could bias the fit and thereby distort the extracted rate constant inside the cavity.

In addition to direct exponential fitting, the simulated reaction traces were also analyzed using a standard linearization approach used in chemical kinetics. Specifically, we computed $\ln[\text{Ext}_\infty - \text{Ext}(t)]$ and fitted the resulting trace with a linear function. For most data sets, the value of $\text{Ext}_\infty$ used in this transformation was obtained from the preceding exponential fits. When Eq.~\ref{eq:ExpSimple} is used, $\text{Ext}_\infty$ is one of the two fit parameters. When Eq.~\ref{eq:ExpFull} is applied, $\text{Ext}_\infty$ equals $\text{Ext}_0 + \text{Ext}_\textrm{bck}$.

Complete details of the simulations reproducing previously published experiments are provided in the Supporting Information. The SI also contains a full description of our own VSC chemistry experiments, including the procedures and analyses related to the observed cavity contraction.

\section{Results and Discussion}

\subsection{Rate in an ideal cavity}

\begin{figure*}[!t]
    \centering    \includegraphics[width=\linewidth]{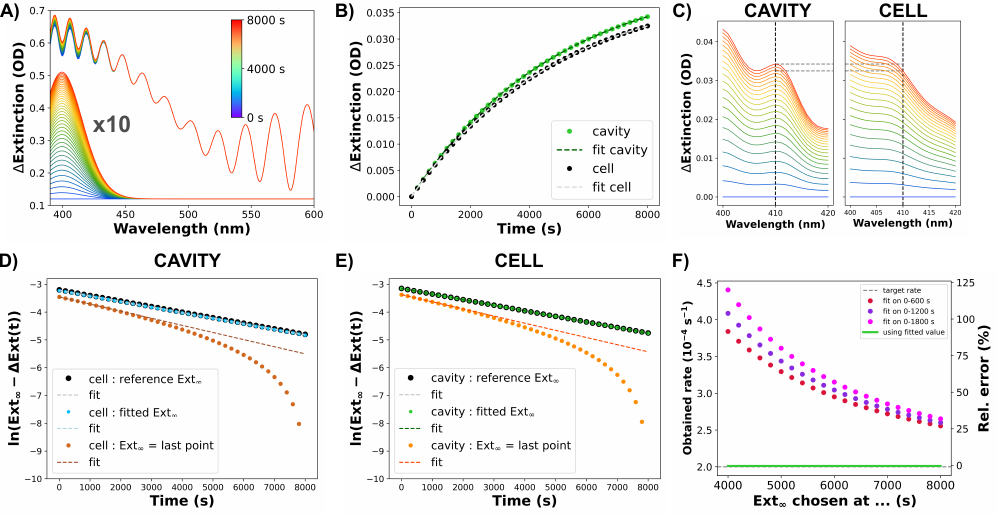}
    \caption{
    \textbf{First order kinetics analysis in an ideal VSC cavity. }  
  \textbf{A)} Simulated UV/Vis extinction spectra of ideal Fabry-P\'erot cavities with perfectly parallel 10 nm thick Au mirrors held at a constant distance of 5 $\mu$m together with the extinction spectra of the absorbing layer between the mirrors (multiplied 10-fold for clarity). The gaussian extinction of the 5 $\mu$m thick absorption layer increases exponentially with constant rate $k_\mathrm{reaction}$ over time (dark blue $\rightarrow$ dark red). Extinction parameters: peak = 400 nm, $\sigma_\mathrm{gauss} = 1000$ cm$^{-1}$, $\tilde{a} = 1.2\times 10^{-4}$, $k_\mathrm{reaction} = 2\times 10^{-4}$ s$^{-1}$. 
  \textbf{B)} Analysis of reaction time trace, i.e., extinction at 410 nm as a function of time for both cavity (green) and cell (no mirrors - black) with exponential fits. 
  \textbf{C)} Comparison of the differential extinction spectra over time of an ideal cavity (left) and cell (right). The double dashed lines indicates where $\Delta\text{Ext} (8000~\text{s})$ is in the cavity and cell, respectively.
  \textbf{D)}  Analysis of linearized reaction time traces for the cell measurement with linear fits. The three traces are calculated with an $\text{Ext}_\infty$ derived by direct computation (black), from an exponential fitting (blue) or selection of the maximum observed extinction (brown).
  \textbf{E)} Analysis of linearized reaction time traces of a cavity setup with linear fits. The three traces are calculated with an $\text{Ext}_\infty$ derived by direct computation (black), from exponential fitting (green) or selection of the maximum observed extinction (orange).
  \textbf{F)} Dependence of the fitted reaction rate, obtained from the linearized trace analysis using the maximum observed extinction as $\text{Ext}_\infty$, on the time a hypothetical experiment was stopped as well as on the size of the fit window used. }
  \label{fig:1}
\end{figure*}

We begin by examining the simplest scenario: an ideal Fabry–P\'{e}rot cavity formed by two perfectly parallel mirrors separated by an absorbing layer. The layer contains a Gaussian-broadened absorption band centered at 400 nm, representing a \textit{para}-nitrophenolate-type (PNP) chromophore (Figs.~\ref{fig:1} and S1). This model is chosen to emulate previous VSC catalysis experiments on PNPA, as well as our own ongoing studies of PNPA hydrolysis under VSC conditions.

To simulate a (pseudo-)first-order chemical reaction with a fixed rate constant of $k_\mathrm{{reaction}} = 2.0\times 10^{-4} \ \text{s}^{-1}$, we scale the absorption intensity of the chromophore over time by modulating the dielectric function of the absorbing layer (see Methods for details). This approach produces a time-dependent increase of the cavity extinction with the expected first-order kinetics, as shown in Figs.~\ref{fig:1}A and B.

Figure \ref{fig:1}A shows the UV/Vis extinction spectra of the ideal cavity at various time points, together with the extinction of the absorbing layer (scaled for clarity) obtained from its time-dependent dielectric function (full and zoomed versions are provided in Figures S1A and B). A gradual increase in extinction in the 400–450~nm region is visible, although its amplitude remains smaller than the modulation introduced by the cavity fringes. This behavior closely resembles typical VSC chemistry experiments on PNPA reported by us and others~\cite{JinoGeorge2, WeiXiong}.

Because the cavity width is fixed in this ideal Fabry–P\'{e}rot structure, the reaction progress can be extracted straightforwardly by taking the extinction at 410~nm at $t=0$ and subtracting it from the extinction at t = 0~s. The resulting time-dependent change in extinction is plotted in Figure~\ref{fig:1}B for both the cavity and, for comparison, a reference structure without mirrors. Following conventions in the field, we refer to the latter as the \emph{cell}, representing a simple short path-length transmission geometry.

To extract the reaction rates, the differential extinction traces (Figure~\ref{fig:1}B) were analyzed by direct exponential fitting (Equation \ref{eq:ExpSimple}). This procedure yields rate constants of 2.0017 for the cavity and 2.0122$\times 10^{-4} \ \text{s}^{-1}$ for the cell, with goodness-of-fit values in the range $\mathrm{0.999 999  < r^2 < 1}$. The corresponding residuals are shown in Figures S1D and S1E, and the numerical stability of the cavity fit is confirmed in Figure S1F.

In practice, reaction rates are often obtained by fitting the linearized form of the differential extinction, $\ln[\text{Ext}_\infty-\Delta\text{Ext(t)}]$.  Because this linearization requires a value for $\text{Ext}_\infty$, the extinction at the end of the measurement is commonly used, meaning that $\text{Ext}_\infty$ is no longer treated as a fitting parameter. As shown in Figure~\ref{fig:1}D for the cell and Figures~\ref{fig:1}E and S1I for the cavity, this procedure can introduce substantial deviations from linearity if the experiment is terminated too early. Nevertheless, to extract a rate constant, the fit is typically restricted to the early-time portion where $\mathrm{\ln[\text{Ext}_\infty-\Delta\text{Ext(t)}]}$ still appears linear. However, as demonstrated in Figure~\ref{fig:1}F, the length of the selected fitting interval has a strong influence on the extracted rate constant and consistently leads to an overestimation. Note that $\text{Ext}_\infty$ differs between the cell and the cavity, as illustrated in Figure~\ref{fig:1}C, and for the cavity, also depends on whether the monitored wavelength lies at a fringe maximum or minimum. However, this variation does not affect the extracted rate constant, as treating $\text{Ext}_\infty$ as a free fitting parameter yields identical rates in all cases.


\subsection{Cavity contraction}

Reliable analysis of (pseudo-)first-order kinetics is straightforward in an ideal Fabry–Pérot cavity in which the mirror separation remains constant throughout the measurement. In practice, however, this condition is rarely met in liquid-phase VSC experiments. Most studies to date have employed microfluidic Fabry–Pérot cavities — essentially modified spectroscopy cells or their derivatives~\cite{EbbesenScience, Hirai, JinoGeorge1, JinoGeorge2, WeiXiong, Vergauwe, SimpkinsScience} (see also the tutorial video from the Ebbesen lab at {https://seafile.unistra.fr/d/7bb78e5a4607424f94b5/}). Although more advanced designs have been proposed~\cite{newVSCcell, Moran2024, EbbesenNMR, Weichman2025}, including devices with fixed cavity width, conventional microfluidic architectures remain the most widely used.
Such cavities can exhibit several types of non-ideal behavior. One is spatial non-uniformity of the cavity width, recently characterized in detail by Michon and Simpkins~\cite{Simpkins}. In our own experiments, we observed a second type of deviation: a gradual contraction of the cavity during reaction monitoring.

\begin{figure}[!t]
    \centering    \includegraphics[width=0.715\linewidth]{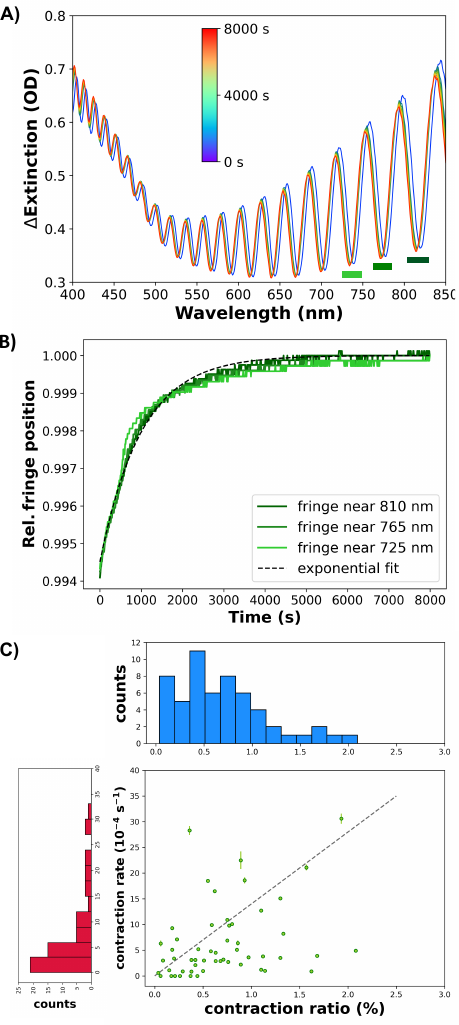}
     \vspace{-2mm}
    \caption{\textbf{Cavity contraction during VSC chemistry experiments. }
  \textbf{A)} Example of a VSC Fabry-P\'erot cavity undergoing cavity contraction after injection of a reaction mixture with PNPA and TABF in methanol. 
  \textbf{B)} Interpolated positions of the three lower energy (longest wavelength) fringes as a function of time. The rate constant of the exponential fit (black dashed line) is 9.9$\times 10^{-4} \text{s}^{-1}$.
  \textbf{C)} Correlation plot between contraction rate and contraction ratio. The dashed line denotes the plot's diagonal. Vertical green bars denote the fitting error of the contraction rate (which for many measurements is too small to be visible on this plot). 
  }
  \label{fig:2}
\end{figure}

Figure~\ref{fig:2}A shows representative unprocessed extinction spectra recorded every 1000~s from a VSC cavity following injection of a reaction mixture containing PNPA and TBAF in methanol. The position of the fringes in these spectra is determined by the residual transmission-mode structure of the cavity. As evident from Figure~\ref{fig:2}A, the spectra do not overlap: the fringes shift systematically toward shorter wavelengths (higher energies) over time (compare, for example, the blue spectrum at $t=0$~s with the red spectrum at $t=8000$~s). This drift is also visible in the zoomed regions shown in Figures~S2A and S2B (400–440~nm and 750–850~nm, respectively).
In a Fabry–P\'{e}rot cavity, the resonance frequencies of the transmission modes vary inversely with the (effective) cavity width. The observed blue-shift therefore indicates a gradual reduction of the cavity width during the measurement. A change in the refractive index of the medium can be ruled out: the reaction converts only the solute PNPA (20~mM) without affecting the solvent methanol ($\sim 30$~M), and the fringe shifts do not correlate with the increasing absorption of the reaction product PNP in the 400–450~nm region. Importantly, this behavior is not unique to this particular cavity. We observed cavity contraction to varying degrees in every cavity sample we measured (see Figure~S3 for 15 additional examples).

The widespread occurrence of cavity contraction can be traced to the way the reaction mixture is introduced into the cavity. Before the monitoring begins, the solution must be pushed through the microfluidic Fabry-F\'{e}rot cell. Under laminar flow conditions, for pressure-driven flow in a rectangular channel, the pressure drop scales as $l^{-3}$ with the smallest channel dimension~\cite{flowphysics}. For microfluidic VSC cavities with spacer thicknesses of only a few micrometers, this scaling implies extremely large pressure drops (on the order of $10^{14}$ Pa) across the channel length. Such pressures cause the cavity assembly to deform elastically and expand slightly. Once the flow stops, the cavity relaxes towards its original geometry, i.e., it contracts over time~\cite{StoppingFlow}. Previous UV/Vis experiments in VSC chemistry typically employed large slit widths, resulting in limited spectral resolution. This smoothing of the spectral fringes masks the visibility of the fringe drift but does not eliminate the underlying cavity contraction.

In the presence of cavity contraction, changes in extinction in the 400–450 nm region no longer reflect only the increasing PNP concentration; they also include contributions from the time-dependent shift of the cavity fringes. It is therefore necessary to model this behavior to validate the extraction of (pseudo)first-order kinetics under such non-ideal conditions using TMM simulations. 

To obtain a phenomenological description of the contraction, we extracted the relative positions of the three lowest-energy fringes over time for each cavity measurement, normalizing the final position to 1, i.e., $\delta\nu_\text{fringe}(t) = [\nu_\text{fringe}(t)-\nu_\text{fringe}(t=8000~\text{s})]/\nu_\text{fringe}(t=8000~\text{s})$. Each resulting time trace was then fitted with an exponential function to quantify the contraction rate, while the contraction ratio was defined as the relative change in fringe frequency between $t=0$ and $t = 8000$~s). The results for the spectra in Figure~\ref{fig:2}A are shown in Figure~\ref{fig:2}B: this cavity undergoes a total relative contraction of approximately 0.6\%, and its temporal evolution is well captured by a single exponential relaxation with a rate of 9.9$\times 10^{-4} \ \text{s}^{-1}$.

This behavior is representative of the majority of the cavities we measured (57 in total; 15 additional examples appear in Figure S3), although the contraction ratios and rates vary from sample to sample. A few cavities exhibit more irregular fringe shifts (\text{e.g.}, Figure~S3E and S3I), and one cavity (Figure S3D) requires two exponentials for distinct time windows. Nevertheless, single-exponential relaxation provides a generally useful description for characterizing contraction in TMM simulations and guiding the relevant parameter range. Figure~\ref{fig:2}C summarizes the analysis for all measured cavities, showing histograms of contraction ratios and rates, along with a correlation plot between the two quantities. Although both parameters span substantial ranges, their correlation is negligible, as indicated by a Pearson coefficient of 0.13.

Having established both a model and quantitative measures of cavity contraction, we next simulate a (pseudo-)first-order reaction occurring in a homogeneous Fabry–P\'{e}rot cavity whose width relaxes exponentially over time. In this framework, the cavity is defined by a single width parameter that changes continuously according to the measured relaxation behavior. Practically, this corresponds to performing a series of TMM simulations of a Fabry–P\'{e}rot structure (as described above), each at a different time point with two time-dependent inputs: (i) the dielectric function of the medium layer, which incorporates the exponentially increasing absorption at 400 nm due to the reaction, and (ii) the cavity width, which models the simultaneous contraction (see Figure S4).


\begin{figure*}[!hbt]
    \centering    \includegraphics[width=\linewidth]{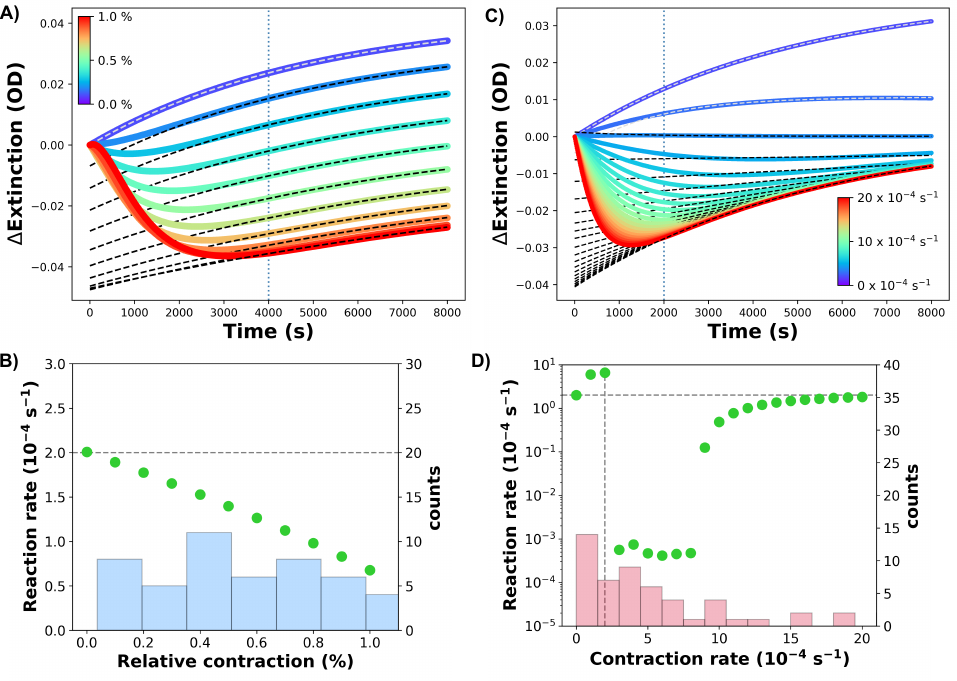}
    \caption{
     \textbf{Effect of cavity contraction on reaction rate determination.} 
     \textbf{A)} Reaction time traces (relative extinction at 410 nm) for cavities undergoing different degrees of contraction (colored curves). The contraction rate is fixed at $9\times10^{-4}\text{s}^{-1}$, consistent with experimental observations.  Dashed line lines show exponential fits to the tail region of each trace (4000-8000 s, indicated by the vertical line).
     \textbf{B)} Extracted rate constants (green dots) as a function of increasing relative cavity contraction ($\delta\nu_\text{fringe}$). The input reaction rate of $2\times 10^{-4} \text{s}^{-1}$ is indicated by the horizontal dashed line. A histogram of experimentally observed relative contractions is shown light blue.
     \textbf{C)} Reaction time traces for cavities contracting at different rates (colored curves), with the total contraction fixed to 0.5 \%. Dashed lines show exponential fits to the 2000-8000 s window, again indicated by a vertical line.
     \textbf{D)} Extracted rate constants (green dots) as a function of increasing contraction rate. The input reaction rate of $2\times 10^{-4} \text{s}^{-1}$ is indicated by the horizontal dashed line. A histogram of experimentally observed contraction rates is shown in light red. 
     Simulation parameters: Spatially uniform Fabry-P\'erot cavity exhibiting an exponential decrease in cavity width towards a limiting value of 5 $\mu m$. The medium layer displays an exponentially increasing absorption with $k_\mathrm{reaction} = 2\times 10^{-4}\text{s}^{-1}$. Monitoring wavelength: 410 nm.
  }
  \label{fig:3}
\end{figure*}


We begin by modeling the UV/Vis spectra of cavities contracting from different initial widths while maintaining a fixed contraction rate of 9$\times 10^{-4} \ \text{s}^{-1}$. The resulting reaction time traces (i.e., the differential extinction at 410 nm), are shown in Figure~\ref{fig:3}A. It is immediately apparent that, if left uncorrected, cavity contraction has a dramatic effect on the apparent changes in the extinction at 410~nm.  $\Delta\text{Ext}$ now exhibits a biphasic evolution: an initial decrease (which can even lead to negative values at higher contraction levels), followed at later times by a return to exponential growth. Consequently, fitting the early portion of these traces invariably yields severely erroneous results, including negative reaction rates, for both direct exponential fitting and linearized analysis (see Figures~\ref{fig:3} and S6). Fitting only the late-time segments of the traces (4000-8000 s; fits with $\mathrm{0.999 < r^2 < 1}$; see residuals in Figure~S6) produces more stable estimates, but these systematically underestimate the true reaction rate as the relative contraction increases (Figure \ref{fig:3}C).

The effects of increasing contraction rates on reaction rate estimation are more complex and potentially more dramatic (Figure \ref{fig:3}B and D). At very low contraction rates, the extracted reaction rates can be overestimated by up to factor of three. At moderate contraction rates (2-8$\times 10^{-4} \ \text{s}^{-1}$), the apparent reaction rates are instead underestimated -- by as much as three orders of magnitude -- because the cavity fringe shifts suppress the extinction increase at 410~nm. However, when the contraction rate significantly exceeds the reaction rate, contraction has little to no influence on the extracted rate constants. In this regime, the cavity width has already reached its equilibrium value prior to the kinetic measurement, such that the UV/Vis signal directly reflects the evolution of product concentration. As a result, the reaction rate can be accurately determined. As indicated by the histogram in Figure~\ref{fig:3}D, such high contraction rates are indeed observed experimentally.

\subsection{Mitigating cavity contraction artifacts}

Given the substantial impact of cavity contraction and the resulting fringe shifts on reaction-rate estimation, we next examine whether standard data-processing techniques can mitigate these artifacts. We consider three commonly used approaches: second-order Savitzky-Golay filtering, rolling average filtering and spectral integration. An alternative strategy would be to increase the bandwidth of the spectrometer’s probe beam by widening the diffraction-grating slit~\cite{Kripa, Yantao}. While this approach effectively smooths cavity fringes without post-processing, it eliminates a key diagnostic of Fabry–P\'{e}rot stability and therefore removes an important indicator of cavity deformation. Moreover, as noted by Michon and Simpkins, spectral smoothing prior to to conversion from transmission to extinction can introduce artificial peak shifts~\cite{Simpkins}. For these reasons, post-processing approaches provide a more controlled and transparent means of handling cavity fringes and their temporal drift.

\begin{figure*}[!bt]
    \centering    \includegraphics[width=\linewidth]{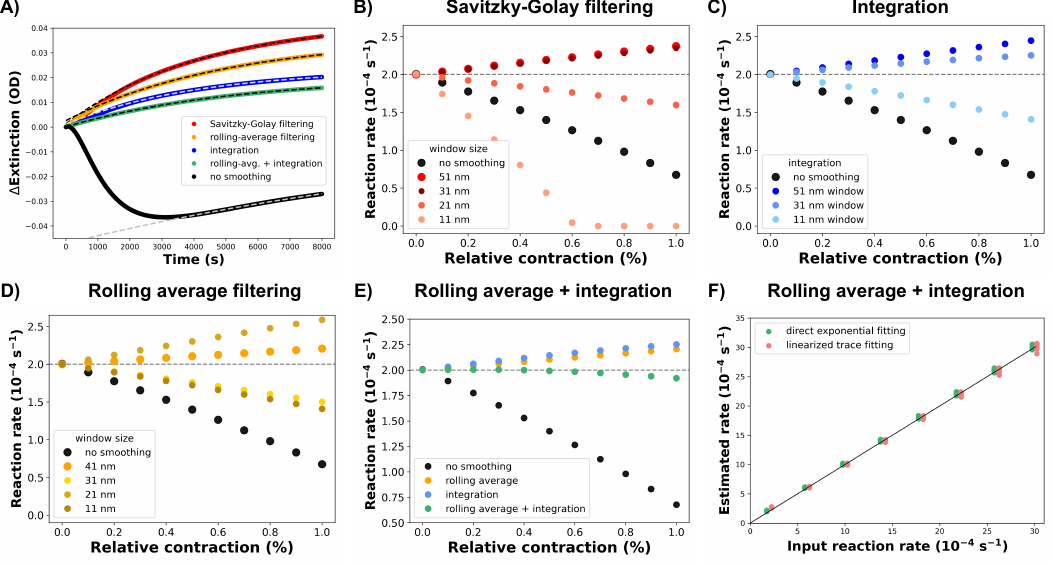}
    \caption{
    \textbf{Mitigation of cavity contraction effects by data smoothing and spectral integration.}
    \textbf{A)} Reaction time traces of a cavity undergoing a 1\% contraction at a rate of $9\times10^{-4}\ \text{s}^{-1}$, extracted after applying no filtering (black curve), Savitzky–Golay filtering (red; 2nd order, 51 nm window), rolling-average filtering (orange; 41 nm window), or spectral integration (blue; 400–450 nm window).
    \textbf{B–D)} Estimated reaction rates for cavities exhibiting different total relative contractions (0.0–1.0\%), while keeping the contraction rate fixed, using (B) Savitzky–Golay filtering, (C) spectral integration, and (D) rolling-average filtering, each with multiple window sizes.
    \textbf{E)} Estimated reaction rates for cavities with different total relative contractions (0.0–1.0\%), with the contraction rate fixed, when applying a combined procedure of rolling-average filtering followed by spectral integration, compared to each technique alone or no smoothing.
    \textbf{F)} Correlation between input and estimated reaction rates for the combined (rolling-average + integration) approach. Each point corresponds to a simulation with a specific set of contraction parameters (relative contraction and rate). Simulation parameters: Spatially uniform Fabry–P\'{e}rot cavity exhibiting an exponential decrease in cavity width toward a limiting value of 5~$\mu$m. The medium layer displays an exponentially increasing absorption with a rate constant $k_\mathrm{reaction} = 2\times10^{-4}\ \text{s}^{-1}$. Monitoring wavelength: 410~nm. Standard fitting window: 2000–8000~s (individual techniques); 0–8000~s (combined approach).
    }
  \label{fig:4}
\end{figure*}

To illustrate the potential of these mitigation strategies, we consider a cavity undergoing a 1\% contraction at a rate of $9\times 10^{-4} \ \text{s}^{-1}$, an extreme case shown in Figure~\ref{fig:3}A. Applying each of the three smoothing techniques to the simulated spectra prior to $\Delta\text{Ext}$ extraction yields the red, orange and blue curves in Figure \ref{fig:4}A. These processed traces differ markedly from the unsmoothed black trace: most of the distortion in the first 3000~s is suppressed, although the early portion still deviates slightly from ideal exponential behaviour. Exponential fitting of the 2000-8000~s interval gives extracted rates of
\begin{itemize}
    \item 2.3745$\times 10^{-4} \ \text{s}^{-1}$ (18 $\%$ rel. error) for Savitzky-Golay filtering 
    \item 2.2067$\times 10^{-4} \ \text{s}^{-1}$ (10 $\%$ rel. error) for rolling-average filtering 
    \item 2.4458$\times 10^{-4} \ \text{s}^{-1}$ (22 $\%$ rel. error) for spectral integration.
\end{itemize}
All three represent a substantial improvement compared to the unsmoothed data, which yield an apparent rate of 0.6768$\times 10^{-4} \ \text{s}^{-1}$, corresponding to a 68\% relative error.

Looking more closely at each technique individually, we can summarize the following behavior when testing against increasing levels of relative contraction:
\begin{itemize}
    \item \textbf{Savitzky–Golay filtering:} A window of 21~nm (21 data points) is required to produce noticeable improvement, and even under high contraction levels the relative error does not fall below $\sim$16\% (Figure~\ref{fig:4}B).

    \item \textbf{Spectral integration:} Integrating over a 31~nm spectral region (31 data points) yields the best performance, with relative errors up to $\sim$23\% and no further improvement for larger windows (Figure~\ref{fig:4}C).

    \item \textbf{Rolling-average filtering:} The largest tested window size, 41~nm, produces the smallest relative error (Figure~\ref{fig:4}D). For the limiting case of a 1\% contraction, the relative error is approximately 10\%, the lowest among the three methods.
\end{itemize}
When increasing the contraction \emph{rate} instead of the contraction \emph{magnitude}, similar trends are observed. For all three techniques, the estimation errors are largest at moderate contraction rates (2-8$\times 10^{-4} \ \text{s}^{-1}$) and converge to zero as the contraction rate becomes higher.


Opting for linearized trace fitting instead of direct exponential fitting does not change the overall picture, provided that the linearized traces are computed using fitted $\text{Ext}_\infty$ values (see Figures S7 and S8). Errors arising from cavity fringe shifts can occur alongside those introduced by an incorrect choice of $\text{Ext}_\infty$ and/or an inappropriate fitting window (see Figs. S9 and S10). This remains true even when smoothing techniques are applied, as these methods cannot fully correct distortions in the early portion of the reaction progress curve -- precisely the region typically used for linearized kinetic analysis. As a result, the extracted reaction rates can be overestimated by several-fold, as illustrated for Savitzky–Golay filtering and spectral integration in Figures S9 and S10. These findings underscore once again the critical importance of using an accurate $\text{Ext}_\infty$ when performing linearized kinetic trace analysis.


Next, we ask whether the performance of the best individual method -- rolling-average filtering -- can be further improved by combining it with spectral integration. Because rolling-average filtering requires selecting a single wavelength, adding an integration step may provide additional robustness. As shown in Figure~\ref{fig:4}E, this is indeed observed: the combined approach reduces the reaction-rate estimation error to $\sim$4\%, whereas applying no smoothing or using either technique alone results in relative errors of 68\%, 10.4\%, and 12.5\%, respectively. When varying the contraction rate, the combined method retrieves reaction rates with a relative error of 2.2\%, compared to -6.9 \% and 8.2 $\%$ for rolling-average filtering or integration alone (Figure S11).

As a final test, we assessed the reliability of the combined rolling-average \emph{plus} integration approach by computing a calibration curve and verifying its linear response to a change in the input reaction rate. To this end, we simulated and analyzed reaction-progress traces for cavities exhibiting every combination of a set of relative contractions $\mathrm{ = \{ 0.1\% , \ 0.6\%, \ 1.0 \ \% \} }$ and contraction rates $\mathrm{ = \{ 1, \ 5, \ 8, \ 14, \ 20\times 10^{-4} \ \text{s}^{-1} \} }$, across a range of different input reaction rates from 2 to 30 $ \times 10^{-4} \ \text{s}^{-1}$. This procedure mimics a series of polaritonic catalysis measurements in which the intrinsic reaction rate varies while each cavity exhibits a different degree of contraction. The results, shown in Figure \ref{fig:4}F, confirm an approximately linear relationship between the input and extracted reaction rates, both when fitting the reaction time trace, $\mathrm{\Delta Ext(t)}$ directly (green circles) and when fitting the linearized trace (orange circles). The relative spread in the extracted values remains similar across the range of input rates (inset Figure S11G). Together, these results demonstrate that the combined approach of rolling average filtering and spectral integration is the most accurate and robust method among those tested here 
for extracting reaction rates from contraction-affected cavity measurements.

\subsection{Cavity width inhomogeneity}

In addition to cavity-width contraction, VSC Fabry-P\'{e}rot cavities can also suffer from cavity width inhomogeneity. Such non-uniformity may arise from mirror misalignment or from bending of the mirrors under pressure when the cavity is assembled and clamped tightly. Michon and Simpkins~\cite{Simpkins} have shown that spectral distortions caused by spatial non-uniformity can lead to an altered effective molar extinction coefficient within the cavity. They further demonstrated that this static inhomogeneity alone can produce the erroneous appearance of modified reaction rates, although their analysis did not include the effects of dynamic contraction.

To examine the combined influence of both factors on apparent reaction rates, we adopt the approach of Michon and Simpkins and average a set of TMM-calculated transmission spectra over a distribution of cavity widths centered around 5~$\mu m$ with $\mathrm{L_{FWHM}}$ 0–110~nm. For each width in the distribution, identical contraction parameters are applied so that the instantaneous distribution of cavity widths remains consistent at all times during the simulated measurement.

\begin{figure*}[bt]
    \centering    \includegraphics[width=\linewidth]{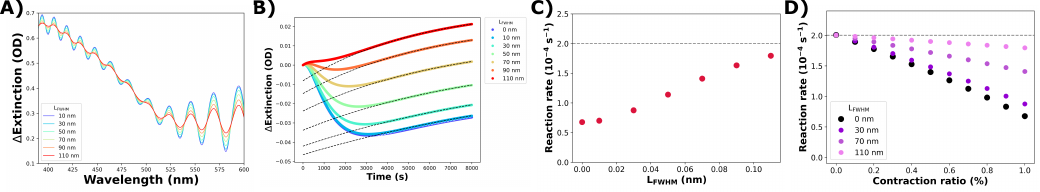}
    \caption{
    \textbf{Effect of cavity width inhomogeneity.} 
  \textbf{A)} Example of simulated spectra at t = 0 s for 6 different levels of gaussian width broadening (as characterized by $\mathrm{L_{FWHM}}$). 
  \textbf{B)} Reaction time traces for the formation of a PNP-like product for 6 contracting cavities with increasing levels of gaussian cavity width inhomogeneity. Dashed lines denote exponential curves fitted to the long-time tails of the curves. Relative contraction = 1.0\% and and contraction rate = $9\times 10^{-4} \ \text{s}^{-1}$.
  \textbf{C)} Reaction rates as obtained from direct exponential fitting to the time traces displayed in panel B). 
  \textbf{D)} Effect of increasing cavity width inhomogeneity on reaction rate determination in cavities experiencing different levels of contraction after injection. 
  Contraction rate = $9\times 10^{-4} \ \text{s}^{-1}$.
    }
  \label{fig:5}
\end{figure*}

The simulations described above produce UV/Vis spectra of the type shown in Figure~\ref{fig:5} (expanded in Figure~S13A), where increasing the FWHM of the cavity-width distribution leads to a progressive fading of the cavity fringes.
This behavior is characteristic of cavity-width inhomogeneity and corresponds both to what we have directly observed during the preparation of VSC cavities and to what Michon and Simpkins previously reported~\cite{Simpkins}. Notably, an $\mathrm{L_{FWHM}}$ of 110~nm -- only 2.2\% of the nominal 5~$\mu m$ cavity width -- is already sufficient to attenuate the fringes almost completely. This highlights the pronounced sensitivity of Fabry–P\'{e}rot cavities to even small variations in mirror spacing.

Examining the complete set of simulated reaction time traces shows that increasing $\mathrm{L_{FWHM}}$ reduces the impact of cavity contraction at early time points (Figure~\ref{fig:5}D), and also decreases the deviation of the extracted reaction rate from the true input value (Figure~\ref{fig:5}C). Systematic modeling of reactions in inhomogeneous cavities with different contraction magnitudes makes this trend even clearer: at large $\mathrm{L_{FWHM}}$ values, the error in the fitted reaction rate grows much more slowly than in cavities with small $\mathrm{L_{FWHM}}$, or in the homogeneous case without width broadening ($\mathrm{L_{FWHM}}$ = 0 nm; Figure~\ref{fig:3}). Varying the contraction rate instead of the contraction ratio yields a similar behavior (Figures~S13C and S13D). 
For the model reaction considered here, cavity-width inhomogeneity affects the simulated spectra in a manner that can partially mask the signatures of cavity contraction, similar to spectral smoothing or increased spectrometer slit width. However, this behavior is specific to the present model system and should not be interpreted as a general cancellation of artifacts. In practice, both effects remain sources of uncertainty and should be characterized, mitigated where possible, and accounted for in the rate analysis.

\subsection{Impact of our findings on previous reports}

We have used TMM simulations to identify four pitfalls that can affect the extraction of reaction rate constants in VSC chemistry experiments based on UV/Vis spectroscopy: (\textit{i}) the choice of the limiting value $\text{Ext}_\infty$, (\textit{ii}) the fitting interval, (\textit{iii}) the cavity contraction, and (\textit{iv}) cavity-width inhomogeneity (the latter already described by Michon and Simpkins~\cite{Simpkins}). Together, these factors can introduce artifacts that affect the reliability of extracted rate constants. 
To assess whether the artifacts identified here could help reconcile discrepancies between four published studies reporting different reaction rates for PNPA hydrolysis in Fabry-P\'{e}rot cavities under VSC, we now evaluate to what extent these artifacts can alter the apparent rate relative to the true underlying rate.

\subsubsection{Ten-fold rate enhancement (Lather \textit{et al.} 2019)}

The first report of a rate enhancement under vibrational strong coupling was published by George and co-workers~\cite{JinoGeorge1}, who observed an apparent tenfold acceleration of PNPA hydrolysis based on UV/Vis monitoring of the reaction product inside an $\sim$18~$\mu m$ Fabry-P\'{e}rot cavity tuned to the C=O stretch vibration of both PNPA and ethyl acetate. From the methodological details provided in the paper, it appears that no spectral smoothing was applied and that the reaction progress was monitored at a single wavelength. Our analysis~(Fig.~S14) shows that, within the parameter ranges we have explored, an apparent reaction acceleration of this magnitude can indeed arise when the cavity contraction occurs on a similar timescale as the reaction and is large in relative terms (exceeding $\sim$~0.8\%). In our own experiments, we have encountered contractions of this magnitude and rate, and therefore consider it plausible that the rate enhancement reported by Lather~\textit{et. al.}~\cite{JinoGeorge1} may be attributable, at least in part, to cavity-contraction effects.

\subsubsection{Three-fold rate enhancement (Singh \textit{et al.} 2023)}
In a follow-up study, George and co-workers~\cite{JinoGeorge2} revised their earlier claim of a tenfold enhancement and instead reported a threefold increase in the hydrolysis rate of PNPA, as well as a similar effect for bis-(2,4-dinitrophenyl) oxalate. Crucially, in this work they applied Savitzky–Golay smoothing to the measured spectra. Our analysis (Fig.~S15) shows that this processing step can substantially reduce the impact of contraction-induced distortions on the extracted kinetic traces.
Only at very low reaction rates ($\mathrm{\times 10^{-3} \ \text{s}^{-1}}$) could contraction-induced artifacts lead to a modest overestimation of the extracted rate—an effect far too small to account for the reported threefold enhancement. Thus, the observations in~\cite{JinoGeorge2} cannot be explained by the contraction artifacts identified here. However, because no details were provided on how $\text{Ext}_\infty$ was determined, we cannot rule out the possibility that part or all of the reported enhancement arises from the use of an incorrect 
$\text{Ext}_\infty$ value.

\subsubsection{No rate modification (Wiesehan and Xiong, 2021)}
The study by Weisehan and Xiong~\cite{WeiXiong} was the first independent attempt to replicate the findings of Lather \textit{et al.}~\cite{JinoGeorge1} and provides a commendably detailed account of both the experimental procedures and the data-analysis methodology. Our analysis shows that their use of spectral integration over a broad wavelength range effectively removes any influence of cavity contraction on the extracted time traces. As a result, the reaction rates they obtain exhibit high fidelity, with relative errors within approximately 3\% (Figure~S16).

\subsubsection{1.8-fold rate enhancement (Verdelli \textit{et al.} 2021)}
The group of G\'{o}mez-Rivas employed both plasmonic nanoparticle metasurfaces and microfluidic Fabry-P\'erot cavities to study PNPA cleavage under VSC~\cite{Jaime}. We do not comment on their metasurface experiments, which were the main focus of their study. Only limited details were provided for the Fabry-P\'erot cavity measurements, preventing us from confidently reproducing their experimental configuration in our simulations. In particular, because essential information on their data-analysis procedure is missing, we do not know whether any smoothing was applied. In light of our finding that, without smoothing, unavoidable cavity contraction can dramatically inflate the apparent reaction rate relative to the true rate, we cannot rule out the possibility that the 1.8-fold rate enhancement reported by Verdelli \textit{et al}. arises from this artifact.

\section{Conclusion}
We have used a controllable TMM framework to identify the challenges and pitfalls inherent in extracting (pseudo-)first-order kinetics from UV/Vis monitoring of reactions inside commonly used Fabry-P\'erot VSC cavities based on modified IR flow cells. We find that reaction rates can be significantly under- or overestimated due cavity-width contraction, which perturbs the residual cavity fringes in the UV/Vis region. Such contraction reflects the mechanical relaxation of the cavity following reactant injection and appears unavoidable given the flow mechanics associated with the small dimensions of typical IR flow cells~\cite{StoppingFlow}. 

The impact of cavity contraction can, however, be mitigated to a large extent by spectral smoothing. 
Among the protocols tested, the most effective procedure for extracting reliable reaction traces from UV/Vis data combines rolling-average smoothing with subsequent spectral integration. It is unfortunate that spectral filtering has not been consistently implemented in prior reports of PNPA cleavage under VSC, as this raises questions about the reliability of some of the published rate enhancements. %
Contraction-induced artifacts can act together with more traditional sources of error in first-order kinetic analysis -- such as incorrect choice of $\text{Ext}_\infty$ or inappropriate fitting windows -- but cross-validating direct and linearized fits can help diagnose these issues. Without appropriate mitigation strategies, studies using microfluidic Fabry–P\'{e}rot cavities for VSC chemistry risk yielding unreliable results.

Beyond analysis methodology, new microfluidic VSC cavity designs may offer improved mechanical stability~\cite{newVSCcell}, and fixed-width cavities have already shown promise for systematic reaction screening under VSC~\cite{EbbesenNMR, Moran2024, Moran2025}, albeit with certain trade-offs. Alternative platforms, such as metasurfaces, avoid the issue of cavity-width contraction altogether~\cite{Jaime, JoelCuSO4}, though they introduce distinct experimental challenges.
Improvements in cavity transmission in the UV/Vis regime may also be beneficial~\cite{Weichman2025}.

More broadly, our results -- together with those of Michon and Simpkins~\cite{Simpkins} -- highlight that probing reactions or material properties inside photonic structures via spectroscopy is considerably more complex than in conventional physical chemistry or materials science settings~\cite{Renken2021}. Yet, because the spectral response of photonic structures can generally be computed using well-established techniques (TMM for planar geometries and FDTD or FEM for other nanophotonic architectures~\cite{FundPhot}), we urge the community to validate their analysis protocols against simulations before drawing conclusions from experimental data.

\subsection*{Acknowledgement}
We would like to thank for Pasi Myllyperkiö and Eero Hulkko for assistance with the spectrometers during the cavity measurements and Heikki H\"akk\"anen for making an adapter mount for this work.

\subsection*{Funding}
RV acknowledges funding from the European Research Executive Agency under the European Union’s Horizon TMA MSCA Postdoctoral Fellowships – European Fellowships action (Grant Agreement No 101068621). JJT acknowledges funding by the Research Council of Finland (Project 363656) and the Finnish Cultural Foundation.



\subsection*{Conflict of interest}
The authors state no conflict of interest.

\subsection*{Data availability statement}
The datasets generated and/or analyzed during the current study are available from the corresponding author upon reasonable request.

\onecolumn
\bibliographystyle{IEEEtran}
\bibliography{bib_VSCanalysis}

\end{document}


\renewcommand\thefigure{S\arabic{figure}}
\renewcommand\baselinestretch{1.1}
\selectfont

\section{\huge Methods}
\subsection{\Large Simulations}
\subsubsection{TMM model}
All simulations of all Fabry-Perot structures were done with the Transfer Matrix Method (TMM). In TMM, a Fabry-P\'{e}rot cavity (taken at a certain time point) is modeled as a stack of layers with a defined thickness and dielectric functions (dependent on photon wavelength): 
\begin{enumerate}
    \item entry medium: air/vacuum
    \item Au mirror: 10 nm
    \item medium layer: 5~\um (limiting value if cavity is contracting; different for simulations of Lather et al.\cite{JinoGeorge1}, Singh et al.\cite{JinoGeorge2} and Weisehan et al.\cite{WeiXiong} - see below)
    \item Au mirror: 10 nm
    \item exit medium: air/vacuum
\end{enumerate}
10 nm is the standard tickness for Au mirrors in VSC chemistry experiments. For each layer and each interface, and for each probe wavelength, a matrix will be constructed. For each probe wavelength, a stack of matrices will be multiplied to one another to yield a final matrix of the simulated system. Standard formalisms for the propagation and TE/TM interface matrix, like described in ref. 4 \cite{FundPhot}, were implemented in Python (Jupyter Notebook). From the final matrix, the TE and TM-polarized transmission spectra, $\mathrm{T_{TE}(\lambda,t)}$ and $\mathrm{T_{TM}(\lambda,t)}$, are obtained (for clarity, these are the intensity transmission coefficients). To consider unpolarized light used in commercial UV/VIS spectrophotometers typically used for VSC chemistry experiments thus far, strictly speaking, the average of these needs to be computed. However, because only normal incidence spectra are considered in this work,  $\mathrm{T_{TE}(\lambda,t)}$ equals $\mathrm{T_{TM}(\lambda,t)}$ and either can be taken as the cavity transmission spectrum $\mathrm{T_{sim}(\lambda,t)}$ and converted to extinction units: 
\[
\text{Ext}(\lambda, t) = -\log_{10}\left( \text{T}_\text{sim}(\lambda, t) \right)
\].

The dielectric function of Au was based on the published values for the complex refractive index obtained from Au thin films by Rosenblatt et al. \cite{Rosenblatt}, previously using by Michon and Simpkins in their simulations of UV/VIS of VSC Fabry-Perot cavities. \cite{Simpkins} 

\subsubsection{Medium layer dielectric constant and reaction simulations}
The dielectric function of the medium layer $\mathrm{\epsilon_{medium}(\omega)}$ consists of a Gauss-Lorentz oscillator model $\mathrm{\epsilon_{absorber}(\omega)}$ with a constant dielectric background $\mathrm{\epsilon_0}$. This models a typical chromophore with a single broad absorption band with a approximately gaussian-like shape, like \textit{para}-nitrophenolate in a (near-)dispersionless organic solvent with refractive index similar to methanol. The implementation of Orosco and Coimbra\cite{Orosco} is chosen as this is consistent with the Kramers-Kronig relations and doesn't involve (numerically expensive) convolution (which the Brendel-Bormann model would require):
\[
\epsilon_\mathrm{medium} = \epsilon_0 + \tilde{a} \cdot\epsilon_{\mathrm{absorber}}(\omega) =\epsilon_0 + \tilde{a}\left[ \frac{ s_w(z_{+}) + s_w(z_{-}) }{\chi_0} \right]
\]
In this construct, $\mathrm{\tilde{a}}$ allows us to control the magnitude of $\mathrm{\epsilon_{absorber}(\omega)}$ contribution and is for most simulations kept at $1.2\times10^{-1}$, while $\chi_0$ is a normalization constant and $\mathrm{s_w(z)}$ is a dimensionless shape function given by:
\[
\chi_0 = -4\sqrt{\pi}D\left( \frac{-\omega}{\sqrt{2}\sigma} \right)
\]

\[
s_w = i\pi\cdot w(z) +  e^{-z^2} \left( \ln{z} + \ln{ \frac{-\bar{z}}{ \vert z \vert^2 } } - i\pi \right) 
\]

with $\mathrm{w(z)}$ and $\mathrm{D(x)}$ resp. the Faddeeva function and the Dawson's function and $\mathrm{z_\pm}$ defined as:
\[
z_+ = \left( \frac{\alpha(\omega) + \omega}{\sqrt{2}\sigma}  \right) 
\]
\[
z_- = \left( \frac{\alpha(\omega) - \omega}{\sqrt{2}\sigma}  \right) 
\]
\[
\alpha(\omega) = \sqrt{\omega ^2 +i\omega\Gamma}
\]
where $\omega$, $\Gamma$ and $\sigma$ are the photon frequency, the Lorentz distribution width and Gaussian distribution width. Note that in the above, signs were chosen to be consistent with the sign convention of our TMM implementation: $\mathrm{\tilde{n}_{medium} = \sqrt{\epsilon_{medium}}}$ will have $\mathrm{Im(\tilde{n}_{medium}) > 0}$ and a proper orientation of the sigmoidal profile of $\mathrm{Re(\tilde{n}_{medium})}$. Using this with different TMM packages may require adapting this. 

In this work, we focus on simulating VSC chemistry experiments in which \textit{para}-nitrophenolate acetate is converted to \textit{para}-nitrophenolate for reason stated in the mean text.\cite{JinoGeorge1, JinoGeorge2, WeiXiong} To that end, we make the following choices for the parameters of the dielectric function of the medium layer, which are held fixed throughout this work:
\begin{itemize}
    \item $\epsilon_0 = (1.30)^2$ For reference, n(MeOH) is approx. 1.31 across the visible spectrum; for most organic solvents n(Vis region) lies between 1.3 - 1.55.
    \item $\lambda_{peak} = 400~nm $
    \item $\Gamma = 1~cm^{-1}$
    \item $\sigma = 1000~cm^{-1}$ which results in a peak FWHM of approx. 25 nm, similar to \textit{para}-nitrophenolate.
    \item $\tilde{\alpha} = 1.2\times10^{-4}$ which results in an extinction of about 0.03-0.035 at 400 nm in both a cell and cavity geometry (\ie without or with mirrors present), typical for a kinetics experiments done with short pathlengths or low concentrations of \textit{para}-nitrophenolate.
\end{itemize}
The extinction spectra obtained with these parameters are shown in Figures 1A, S1A and S1B. The refractive index dispersion due to these absorption amounts to at most $\pm 0.0006$ RI units.

To simulate a (pseudo-)first order reaction with pro-chromophoric substance, a scaling coefficient for the amplitude parameter $\tilde{a}$ of $\epsilon_{\text{GL}}(\lambda)$ with values between 0 and 1 is calculated for different time points:
\[
c(t) = 1 - e^{ - \text{k}_\mathrm{reaction}.t }
\]
For each time point, we run a TMM simulation during which the scaling coefficient is multiplied to $\mathrm{\tilde{a}}$ and passed to a routine for calculating $\mathrm{\epsilon_{absorber}(\omega)}$. 
In almost all cases, time points are chosen to run from 0 to 8000~s at 10~s intervals and spectra where computed from 390 to 850~nm with a sampling resolution of 1~nm. This matches VSC chemistry experiments we have undertaken with PNPA and TBAF. $\mathrm{k_{reaction}}$ is set to $2\times10^{-4}$ throughout this work, save for the case of {\em e.g.}~Figures 4F and S14 where the precision of reaction rate determination is tested for a wide range of $\mathrm{k_{reaction}}$ values, and, when simulating previous works on PNPA cleavage (see Figs.~S14, S15, S16). 

\subsubsection{Reactions outside cavity}
To calculate the time-dependent UV/VIS extinction spectra of a reaction taking place in standard transmission measurement cell as reference geometry, a TMM calculation is performed on an optical layer stack lacking the Au mirrors: air ($n=1$) / medium layer / air ($n=1$). Incorporation of the reaction and the calculation of the cell extinction spectra from the computed $\mathrm{T_\text{TTM}(\lambda,t)}$ was the same as above for the cavity.

\subsubsection{Cavity inhomogeneity}
In order to incorporate lateral cavity width inhomogeneity, we followed the approach of Michon and Simpkins as described in \cite{Simpkins}. 
Briefly, this method considers the case where a typical VSC Fabry-Perot cavities is probed over an area with a non-uniform cavity width. In practice, this manifests itself by colored Newton rings being visible when looking at/through the cavity. Michon and Simpkins illustrate this in detail and we have encountered this routinely when working with microfluidic VSC cavities. Most of the time, the Newton rings will be in cross-pattern resembling the contours of a saddle surface. In that case, there will gradients of cavity width radiating outwards from the center towards both higher and lower cavity width depending on the direction. When considering the entire probe surface as a collection of probe points for a light beam of a UV/VIS spectrometer, it is reasonable to approximate the distribution of cavity widths by a gaussian distribution characterized by a length $\mathrm{L_{FWHM}}$ proportional to the gaussian distribution standard deviation:
\[
\sigma_\mathrm{FWHM} = \frac{L_\mathrm{FWHM} }{2\sqrt{2\ln{2}}} 
\]
The transmission spectrum of a Fabry-Perot structure is  taken as the weighted sum of 17 individual transmission spectra of individual cavities with different width distributed from $\mathrm{L-4.L_{FWHM}}$ to $\mathrm{L+4.L_{FWHM}}$ at $\mathrm{0.5.L_{FWHM}}$ interval with L the central cavity width and with a computed normalized weight:
\[
w(L_i) = \frac{1}{\sqrt{2\pi}\cdot\sigma _\mathrm{FWHM}}\cdot\exp{ \left[ \frac{- ( L-L_i )^2 }{ 2\sigma _\mathrm{FWHM}^2} \right] }
\]

We have validated the above approach for computing spectra of inhomogenously broadened cavities by comparing it with spectra computed as the sum of single-width cavity spectra, for which the cavity widths were randomly generated from a Gaussian distribution (see Fig \ref{fig:S12}). As the number of single-width spectra summed increases, the resulting spectra converge to those obtained by the Michon and Simpkins method. However, a large number of spectra, at least 1024, is needed in order to obtain acceptable convergence and noise level, which comes at the cost of a much longer compute time. Moreover, the reaction progress traces obtained still carry in them a substantial degree of noise (due to the stochastic nature of this computational process). The Michon and Simpkins method therefore seems to us the preferable method. The reason for the appearance of noise is the stochastic nature of this type of modeling versus the deterministic nature of the Michon and Simpkins method. The contributions of lowest and highest cavity width have a very low relative weight (on the order of 0.001) but still impact the final result and thus need to be properly accounted for. This requires a number of samples on the order of 1 / 0.001 = 1000 or more. This is illustrated by Fig. \ref{fig:S12} C, where the noise becomes more or less limited at 1024 and 4096 samples at which a more-or-less stable number of very low or very high cavity widths are incorporated at each time point.

\subsubsection{Data smoothing techniques and reaction time trace fitting}
Three analysis techniques were tested and implemented as follows. 

\textbf{Savitzky-Golay filtering} which involves fitting a polynomial to a local neighborhood of each point of the simulated spectra. Here we chose to utilize Scipy's \texttt{savgol\_filter()} function with a second order polynomial and local fit windows of 51 points (values of 11, 21 and 31 points were also tested). In order to deal with the spectra limits gracefully, the first 10 data points (390-400 nm) of the simulated spectra were omitted and the keyword argument \texttt{mode='mirror'} was passed to extend the input spectrum by mirroring it. The default is \texttt{mode='interp'} and results obtained using this option are similar to using \texttt{mode='mirror'}.

\textbf{Rolling-average filtering} which involves taking a local average for each data point of a simulated spectrum. Michon and Simpkins provide a Python implementation for this filter which we use:
\begin{python}
def moving_average(x_array, y_array, kernel_size = 5):
    J = kernel_size
    kernel = np.ones( 2*J+1 ) / ( 2*J+1 )
    return x_array[J:-J], np.convolve(y_array, kernel, mode='valid')
\end{python}
A smoothing window of 41 data points was standard (window sizes of 11, 21 and 31 were also tested). In the latter two cases, the reaction time trace was taken at the first points of the smoothed spectra (413 and 415 nm resp.). 

\textbf{Integration} which involves integrating a local region of each simulated spectra of 11, 31 or 51 points. This corresponds to the 405-415, 395-425 and 400-450 nm spectral regions. The integration result was divided over the length of the integration window to allow easy comparison with other spectra (raw and filtered).

\textbf{Integration after rolling-average filtering}. After application of a rolling-average filter over a 41 point window (so 20 points on each side of each data point), the resulting extinction spectrum was integrated between 410 and 430 nm.

The progress of the reaction was either inferred from the differential absorption at 410~nm with respect to the absorption at time 0, \ie $~\mathrm{\Delta Ext(410~\text{nm},t) = Ext(410~\text{nm},t) - \\ Ext(410~\text{nm},0~\text{s})}$ (cavity) or from the integrated differential absorption spectrum (cell).

Direct fitting of the time traces of the differential absorption/extinction was done with one of these two exponential model functions: 
\begin{equation} \label{eq:ExpSimple}
    \Delta \text{Ext}(t) = \text{Ext}_\infty \left(  1 - e^{-\text{k}_\textrm{fit} t} \right)
\end{equation}
or
\begin{equation} \label{eq:ExpFull}
    \Delta \text{Ext}(t) = \text{Ext}_0 \left[  1 - e^{ - (t - t_0)\text{k}_\textrm{fit}} \right] + \text{Ext}_\textrm{bck}  
\end{equation} 
where $\mathrm{k_\text{fit}}$ is the fit parameter reflecting the reaction rate; any deviation from $\mathrm{k_{reaction}}$ must thus reflects an artifact. The exponential function in equation~\ref{eq:ExpSimple} is used when examining an ideal, homogeneous cavity (Figures 1 and S1), as there is no time delay and no background extinction in these simulations (typically present during experiments). When cavity contraction is included (see Figure 3, S5 and S6), fitting is performed with an 4-parameter exponential model function \ref{eq:ExpFull} which is essentially an exponential function free to be translated in both the dimensions of time and extinction through the parameters of $t_0$ and $\mathrm{Ext_{bck}}$. This is a fit function which is typically used in analysis of experimental data as, in this case, one will typically encounter measurement time delays (stemming from the delay between mixing the reaction compounds and starting the data collection on the spectrophotometer) and background signals (\eg stray light, uncorrected reflections). To be clear, our modeling assumes neither of these two phenomena to be present (so the reaction starts at t = 0 s with an initial product concentration of 0 M). We employ this fit function here because we want to characterize parts of the reaction time traces affected by contraction artifacts with a method which would normally get used in an experimental setting. The $t_0$ and $\mathrm{Ext_{bck}}$ thus do not possess any physical meaning, but are necessary because the presence of artifacts in the reaction time traces. All fitting was done using the \textit{scipy.optimize.curve\_fit$()$} routine, which performs least-squares curve fitting based on the Levenberg-Marquardt algorithm (no fit constraints were used).~\cite{SciPy-NMeth}

It is standard practice in physical chemistry to analyze (pseudo-)first-order reaction kinetics data by converting this to a linearized form:

\begin{align*}
    \text{Ext}(t) &= \text{Ext}_\infty \left[  1 - \exp{ \left( - \text{k}_\text{reaction}\cdot t  \right)} \right] \Longleftrightarrow \\
\ln{ \left[ \text{Ext}(t) - \text{Ext}_\infty \right]} &=  \ln{ \left[ \text{Ext}_\infty \cdot \exp{ \left( - \text{k}_\text{reaction}\cdot t  \right)} \right]} = \ln{ \left[ \text{Ext}_\infty - \text{k}_\text{reaction}\cdot t \right]} \label{eq:kinlin}
\end{align*}

This practice gained currency at a time when computer-aided data analysis/fitting was not common and stuck with the field ever since. The advantage of this is that:
\begin{enumerate}
    \item verification of first order conditions is straightforward. Both first and second order kinetics data (\textit{i.e.} curves of the order of $\exp{(-t)}$ and $\frac{1}{t}$) look roughly similar to the naked eye when plotted on a normal plot. After conversion to the form stated above, they don't anymore.
    \item only an accurate value of $\mathrm{Ext_\infty}$ is needed, making the need for accurate knowledge of the starting concentrations/extinctions/... unnecessary. In chemical kinetics measurements, there is more than often a time delay between mixing the reagents, which is the true start of the reaction, and the start of data collection. Because reagent conversion occurs at its fastest in the beginning of the reaction, determining accurate starting values can be error-prone.
\end{enumerate}
However, a downside of analysis through linearized time traces is that a precise value of $\mathrm{Ext_\infty}$ must be used. If a correct value is used, $\mathrm{\ln{ \left[ Ext(t) - Ext_\infty \right]}}$ will be linear:
\begin{eqnarray}\nonumber
\ln{ \left[ \text{Ext}_\infty - \text{Ext}(t)   \right]} &=& \ln{ \left\{\text{Ext}_\infty -   \text{Ext}_\infty \left[  1 - \exp{ \left( -\text{k}_\mathrm{reaction}\cdot t \right)}  \right]  \right\}}\nonumber \\ &=& \ln{\text{Ext}_\infty } + \ln{ \left[  - \exp{ \left( - \text{k}_\mathrm{reaction}\cdot t \right)}  \right] }\nonumber \\ &=&  \ln{\text{Ext}_\infty } - \text{k}_\mathrm{reaction}\cdot t \nonumber 
\end{eqnarray}
If a bad value $Ext_\infty^{bad}$ is used, the linearized plot will curve off:
\begin{eqnarray}\nonumber
\ln{ \left[ \text{Ext}_\infty^{bad} - \text{Ext}(t) \right]} 
&=& \ln{ \left\{\text{Ext}_\infty^{bad} - \text{Ext}_\infty \left[  1 - \exp{ \left( -\text{k}_\mathrm{reaction}\cdot t \right)} \right] \right\}}\nonumber \\ 
&=& \ln{ \text{Ext}_\infty } + \ln{ \left[  r - \exp{ \left( -\text{k}_\mathrm{reaction}\cdot t  \right)} \right] } \nonumber
\end{eqnarray}
with $r = \text{Ext}_\infty^{bad}/\text{Ext}_\infty - 1 > 0$. This is illustrated in Figure S1 panel I, where incorrect values for $\mathrm{Ext_\infty}$ are selected on purpose, and in panels C and D of Figure 1. In the latter, two guesses for $\mathrm{Ext_\infty}$ are tested against the 'true' value as determined from a simulation done with the full value of $\tilde{a} = 1.2\times10^{-4}$ (no scaling coefficient applied), which represents a simulate at 'infinite time' so to speak. 

For most of the data in this paper the parameter $\mathrm{Ext_\infty}$ was derived from previous exponential fitting. When fitting with Eq.~\ref{eq:ExpSimple}, $\mathrm{Ext_\infty}$ is just one of the two parameters. When fitting with Eq.~\ref{eq:ExpFull}, $\mathrm{Ext_\infty = Ext_o + Ext_\textrm{bck}}$.

\subsubsection{Simulation of previously published experiments}
To simulate the UV/VIS cavity measurements published by the groups of Jino George and Wei Xiong in their 3 papers~\cite{JinoGeorge1, JinoGeorge2, WeiXiong}, the following general assumptions were made:
\begin{itemize}
    \item the same model for PNP, generated by pseudo-first order reaction from PNPA, was used as for our previous simulated (see above). Parameters of the Gaussian-Lorentzian dielectric function were kept the same: $\lambda_{peak} = 400~nm $, $\sigma = 1000~cm^{-1}$,  $\tilde{\alpha} = 1.2\times10^{-4}$ and $\epsilon_0 = (1.30)^2$.
    \item The 100 nm $\mathrm{SiO_x}$ insulating layers were omitted for simplicity's sake as, in practice, they will not have any discernible influence on the measured UV/VIS spectra as the cavities width are approx. 18-20~\um. 
    \item No information could be found in all 3 papers on the extent of cavity inhomogeneity present in the measured cavities. Because cavity inhomogeneity effectively acts a form of smoothing (see Figure 5 and discussion thereof in the main text), we opted to not include this in our simulations. The simulations thus represent a worst-case scenario.
    \item No information could be found in all 3 papers on the nature of cavity shifting. So, we only examine cavity contractions of the kind we expect from our experiments and also focus on in other parts of this paper, being  contractions of 0 to 1 \% (see Figure S14) or of 0 to 10 \% (see Figure S15) with rates from 0 to 20 $10^{-4} \ s^{-1}$. Our reasoning is that both groups used a flowcell broadly similar (be it commercially sourced) to what we have used.
\end{itemize}

\paragraph{Parameters simulations Lather et al. Angew Chem 2019}
Spectra from 370 to 600 nm (sampled at 1 nm intervals) were computed for 0 to 600~s in 1~s intervals and reaction progress traces were directly taken as the differential extinction at 407~nm without any prior smoothing of the spectra. Although the main text of the article mentions spacer width of approx. 18~\um being using, we instead relied on the table in the Supplementary Information of this paper with free spectral range figures of all the measured cavities when still empty before reaction monitoring (going from 230.5 to 251.3~\cm). From this we calculated the range of measured cavity widths ($\mathrm{l_{cav}=\frac{1}{2.FSR}}$ with $n=1$ for air), which starting at 19.9~\um and going to 21.7~\um. Both values correspond to measurements of OFF resonant cavities, \ie cavities with no VSC of the C=O stretch vibration of ethyl acetate/PNPA. Additionally, we chose to simulate the ON resonant cavity with an FSR of 240.46~\cm and a calculated cavity width of 20.86~\um. As no mention was found regarding how $\mathrm{Ext_\infty}$ was determined, we chose to extrapolate it from exponential fitting. This represents a best case scenario and also avoided the anomalous results (\ie negative fitted rates) we occasionally would encounter when choosing to take $E_\infty$ as the maximum of the simulated trace. The linearized reaction trace was fitted at the first 60 data points, which corresponds to the first 60~s of the measurement.

\paragraph{Parameters simulations Singh et al. Chem Phys Chem 2023}
Spectra from 370 to 600 nm (sampled at 1 nm intervals) were computed for 0 to 1000~s in 10~s intervals. Simulated spectra were smoothed with an 2th order Savitzky-Golay filter (51 nm filtering window) and the reaction progress traces are subsequently taken at 400~nm. We chose to simulated cavity widths of 18~\um and 18.7~\um. The first value is mentioned throughout the article, the second value was computed from only of the only mention of a value for an empty cavity free spectral range in the entire article. This value of 267~\cm was for a cavity near resonance to the C=O stretch vibration of ethyl acetate/PNPA. As no mention was found regarding how $\mathrm{Ext_\infty}$ was determined, we chose to obtain this as the maximum of the simulated reaction trace (no anomalous results were obtained here as Savitzky-Golay filtering suppressed almost all distortions of cavity contraction). The linearized reaction trace was fitted on the first 11 data points, which corresponds to the first 100~s of the measurement.

\paragraph{Parameters simulations Wiesehan et al. J Chem Phys 2021}
Spectra from 370 to 850 nm (sampled at 1 nm intervals) were computed for t = 0 to 600 s at 1.5 s intervals and for t = 600 to 3600 s at 10 s intervals. Smoothing was done by integration over the 375 to 425 nm spectral region. We chose to simulate cavity widths of 18~\um and 19.55~\um. The first value is mentioned throughout the article, the second value is mentioned in the Supplementary Information in the context of a path length calibration between a cell and cavity (otherwise no further mentions of cavity widths or of empty cavity free spectral ranges widths were noted). $\mathrm{Ext_\infty}$ was determined as the maximum of the simulated reaction trace as mentioned in the paper. Fitting linearized reaction traces was done on a window of the first 200 points, which corresponds to the first 300 seconds of the measurement.

\subsection{\Large Cavity contraction experiments}
The experimental cavity contraction data presented in Figures 2 and S3 comes from VSC chemistry experiments with \textit{para}-nitrophenol acetate (PNPA) and tetrabutylammonium fluoride (TBAF) in methanol (MeOH) using a home-made microfluidic VSC Fabry-Perot cavities. These VSC cavities consist out of a pair of insulated mirrors on an optically polished 3 mm thick 1 inch \baf ~IR transparent windows (Crystran Ltd, UK) with is sandwiched in between a spacer cut from polypropylene film of nominally 4\um, but in practice $4.5 - 5.5~\mu m$, thickness (SKU 1000038738 from Goodfellow Cambridge Ltd, UK) . The mirrors were manufactured by thermally evaporating a 10 nm Au layer with a 0,5 nm Ti adhesion layer ($P_{evap} = 1-4\times10^{-8} \ mbar, v_{evap} = 0.2-0.3 \ nm/s$) after plasma cleaning (REI, Oxford instruments Ltd, UK), followed by the growth of an approximately 80 nm \alo insulating layer using ALD (800 cycles at 200° C using trimethyl aluminum and water as precursors; TFS 200 from Beneq OY, Finland). Mirrors and spacer were assembled together in custom made metal holder (steel or brass, each with a rectangular observation port cut out of 18 mm by 11 mm) using O,5 mm thick teflon liners for protection and even pressure distribution. The holder consisted out of a top and bottom part (4 and 10 mm thick respectively) assembled together with 6 M3 screws with 3 spring disks (0.5 mm thick each) equidistantly placed along the edge. Playing with tightness of the screws can result in the appearance of colored Newton rings. In many cases, a spot of a single color of at least 5 mm across can be obtained, which was subsequently selected for spectroscopic interrogation by taping a metal mask with 4 or 5 mm diameter hole to the holder. The cavity was allowed to settle overnight before measurement the next day. The free spectral range increases typically 50 to 100 \cm and is due to relaxation of stresses in the various parts of the cavity assembly. 

Before injection and starting the measurement, each cavity was mounted in a temperature control block mounted on a flip mount equipped with two orthogonal micrometer positioning stages for aligning the cavity to the spectrometer beam. This temperature control block was connected to a refrigerated water recirculator (Thermo Haake C/DC class C10-K10). A 2 mm wide PT100 thermocouple connected to a Keithley 2001 digital mulitmeter (4-point resistance measurement) was inserted to a 2 mm hole in the bottom metal holder for monitoring the temperature of the cavity assembly during each experiment. Before injection and measurement, the cavity was brought to a temperature between 24.9 and 25.1$^\circ$C and allowed to reach thermal equilibrium during a period of at least one hour. During UV/VIS reaction monitoring, the temperature was manually controlled, using the water recirculator, to be in between 24.9 and 25.2$^\circ$C at all times.

The reaction mixture consisted out of 540 $\mu l$ of a 6 M of TBAF stock (abcr GmbH, Germany; tetra-n-butylammonium fluoride trihydrate 97\%, AB112723 , CAS 87749-50-6) in MeOH (analytical grade) mixed with 60 $\mu l$ of 200 mM stock solution of PNPA (Sigma-Aldrich/Merck GmbH, Germany; 4-nitrophenol acetate 98\%, N8130, CAS 830-03-5) in MeOH. Using a 1 ml Luer syringe this mixture was quickly injected into the cavity using 4 mm O.D. plastic tubing and Press-Lock fittings (KJL04-M3 from SMC) with M3 screw threads for easy attachment to threaded holes through the top holder. After injection the syringe was generally left stuck on the input port side, while the tubing on the output port was sealed with a plug of telfon tape and parafilm.

UV/VIS spectra were collected on an Agilent Cary 8454 spectrophotometer (Example shown in Fig. 1A with zooms in Figure S2). This instrument is equipped with both a halogen and deuterium lamp and illuminates the sample with single light beam before analysis with a spectrometer and detection on a CCD array. Full spectra are thus collected in a single shot. For a single experiment, a time series of 800 spectra is collected at 10 s intervals, representing a reaction time interval spanning from 0 s to 8000 s. The light integration time for each spectrum is 0.5 s. A black cloth was used to cover the VSC cavity mounted in the spectrometer and shield it from ambient light.

To characterize the contraction of each measured cavity, the 3 fringes located at the lowest photon energy of each recorded spectra, \textit{i.e.} between 720-840 nm, were tracked. Because the inverse relationship between photon energy and wavelength, the fringes in the red spectral region can be tracked with a greater energy resolution than those in more blue-shifted regions. The cusps, \ie the low extinction dips, rather the high extinction peaks were chosen because the signal-to-noise ratio near the former is usually better than for the latter. To determine the exact position (\ie frequency/energy) of each fringe's cusp, the region around the cusp was selected, interpolated to a 0.1 nm resolution (using \textit{scipy.interpolate.interp1d$($ kind$=$cubic $)$}) and fed to the \textit{scipy.signal.find$\_$peaks$()$} algorithm.~\cite{SciPy-NMeth} The time traces of the fringe frequencies, $\mathrm{\nu_\text{fringe}(t)}$, were normalized the last measured value (\ie at t = 8000 s), setting the final value to 1, \ie, $\mathrm{\delta\nu_\text{fringe}(t) = [\nu_\text{fringe}(t)-\nu_\text{fringe}(t=8000~\text{s})]/\nu_\text{fringe}(t=8000~\text{s})}$. The trace of the cusp of the fringe with the lowest frequency/energy was then fitted with an exponential function (\ref{eq:ExpFull}) to quantify the contraction rate, while the relative contraction was defined as the relative change in fringe frequency between $t=0$ and $t = 8000$~s).

\section{\huge Additional data}
\begin{figure}[h!]
  \includegraphics[scale=0.75]{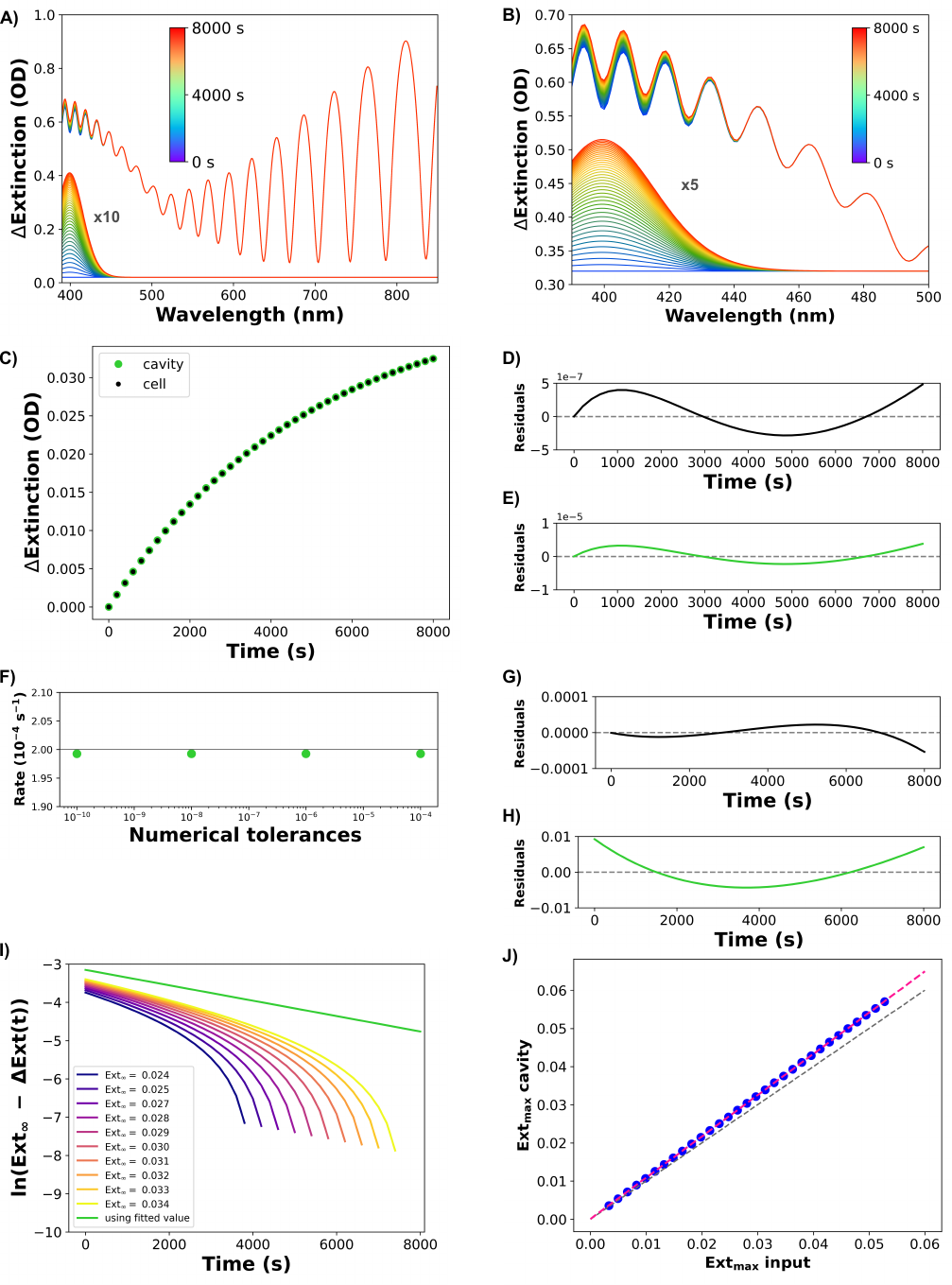}
  \caption{\textbf{First order kinetics analysis in an ideal VSC cavity: additional data.}
  \textbf{A)} Full rendered and \textbf{B)} zoomed in version of Figure 1A. Simulated UV-Vis cavity extinction spectra of an ideal Fabry-Perot VSC cavities together with the extinction spectra of the absorbing layer (multiplied resp. 10- and 5-fold for clarity). Simulations of Fabry-Perot cavity with perfectly parallel Au mirrors (10 nm) held at a constant 5 \um with a medium with exponentially increasing absorption at 400 nm over time with a rate of $\mathrm{k_{reaction}} = 2\times10^{-4} \ \text{s}^{-1})$. 
  \textbf{C)} Rescaled reaction progress curves for cell and cavity simulations shown in Figure 1B.
  \textbf{D)} and \textbf{E)} Residuals of exponential fits to reaction time traces of resp. the cell (black) and cavity (green) in Figure 1B.
  \textbf{F)} Numerical stability of exponential fitting cavity reaction time traces.
  \textbf{G)} and \textbf{H)} Residuals of the linear fits to the linearized reaction time traces calculated with an extrapolated $\mathrm{Ext_\infty}$ value of the cell (black) and cavity (green) in resp. Figure 1C and 1D.
  \textbf{I)} linearized reaction time traces computed with different values for $\mathrm{Ext_\infty}$.
  \textbf{J)} Relationship cell extinction and cavity extinction. Data plot as indigo circles, red and black dashed line denote resp. the linear fit and diagonal of the plot. Different cavities with different medium layer dielectric functions with varying values of \title{a} were simulated to obtain the cavity extinction at 410 nm. For each simulation, the cell extinction was computed directly from the medium layer dielectric function.}
  \label{fig:S1}
\end{figure}

\begin{figure}[ht]
  \includegraphics[width=\textwidth]{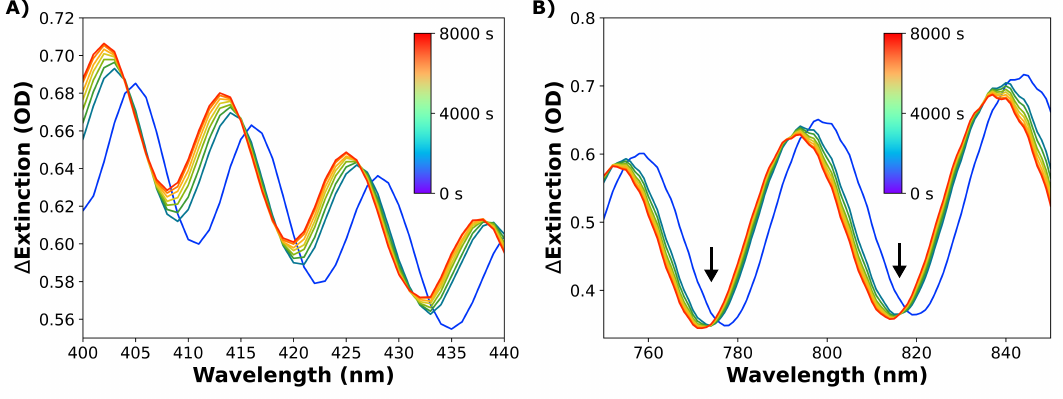}
  \caption{\textbf{Fabry-Perot cavity contraction during a VSC chemistry experiment.} 
  \textbf{A)} and \textbf{B)} Zooms of the spectra displayed in Figure 2A. 
  \textbf{A)} Zoom of the spectral region containing the absorption peak of \textit{para}-nitrophenolate. 
  \textbf{B)} Zoom of the spectral region containing the fringes used for tracking the contraction of the cavity over time. Displayed spectra taken from 0 s to 8000~s at 1000~s intervals.}
  \label{fig:S2}
\end{figure}

\begin{figure}[ht]
  \includegraphics[width=\textwidth]{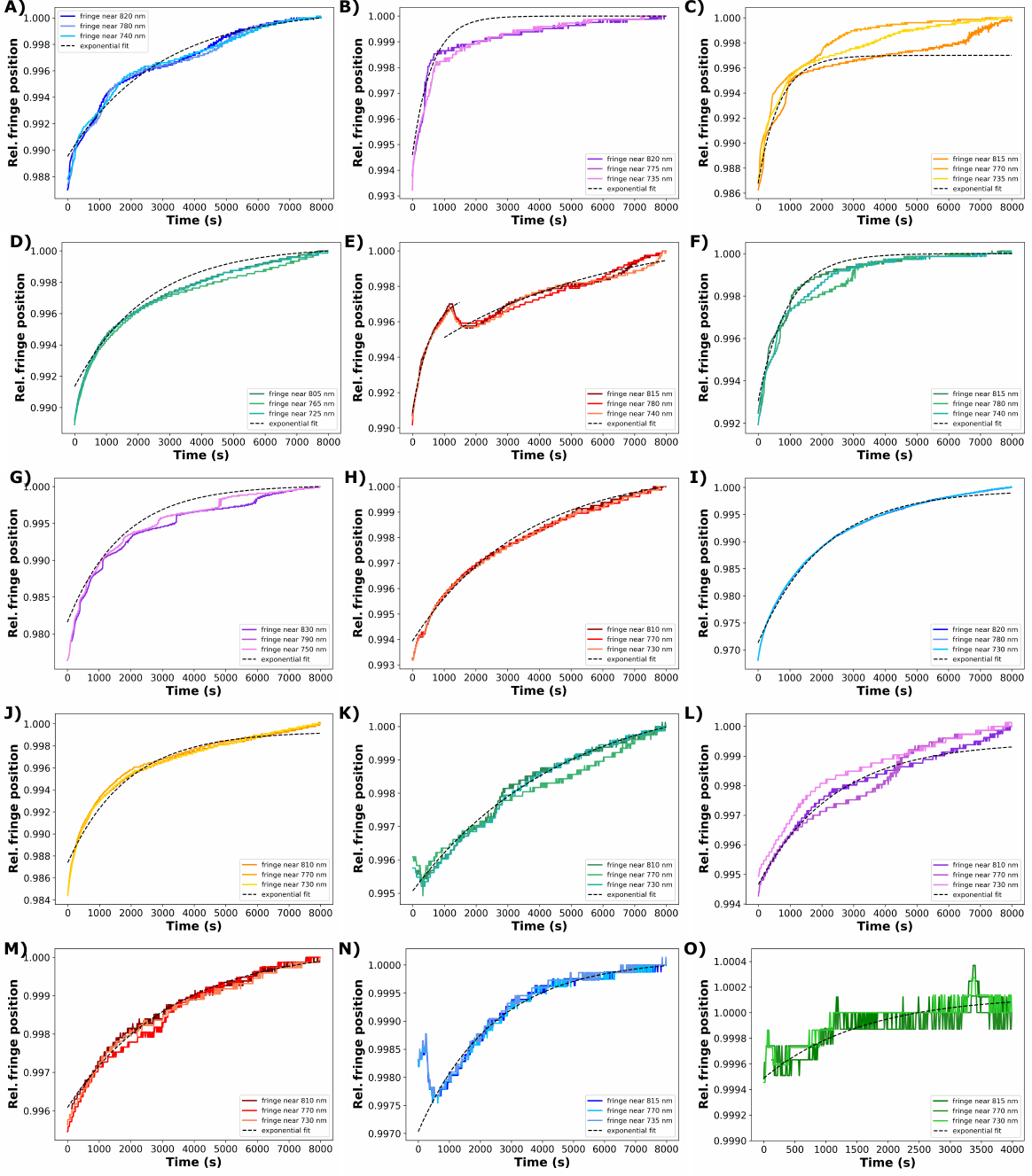}
  \caption{\textbf{Examples of cavity contraction during VSC chemistry experiments.} 
  Time traces of relative energy (spectral position) of the dips of the three lowest energy UV/VIS fringes (see Fig. S2B). Fit of the time trace of the lowest energy cusp plotted as a dashed black line. Data shown for a sample of 15 cavities out of 57 measured and analyzed. 
  }
  \label{fig:S3}
\end{figure}

\begin{figure}[ht]
  \includegraphics[width=\textwidth]{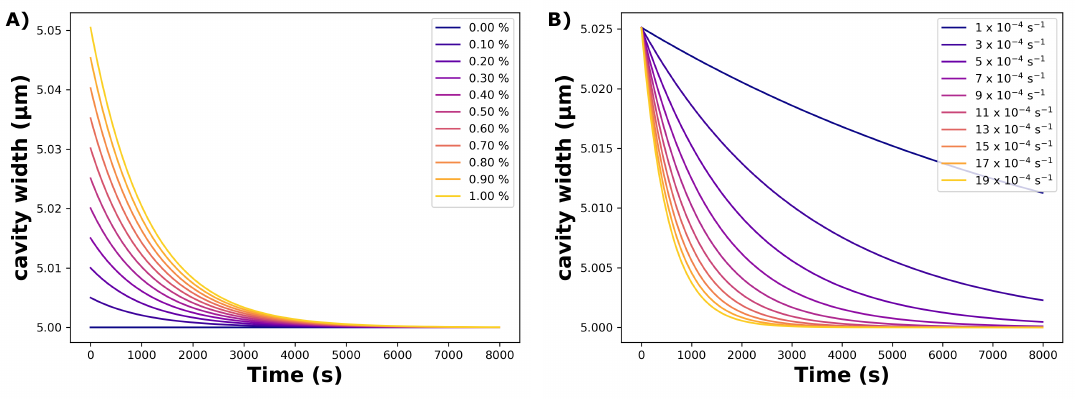}
  \caption{\textbf{Cavity widths vs time used for simulating contracting cavities. }
  A) Cavity widths for simulations with varying relative contractions.  $\mathrm{k_{reaction}} = 9\times10^{-4} \ \text{s}^{-1}$ for all plotted cavity width traces.
  B) Cavity widths for simulations with varying contraction rates. Relative contraction = $0.5 \%$ for all plotted cavity width traces.}
  \label{fig:S4}
\end{figure}

\begin{figure}[ht]
  \includegraphics[width=\textwidth]{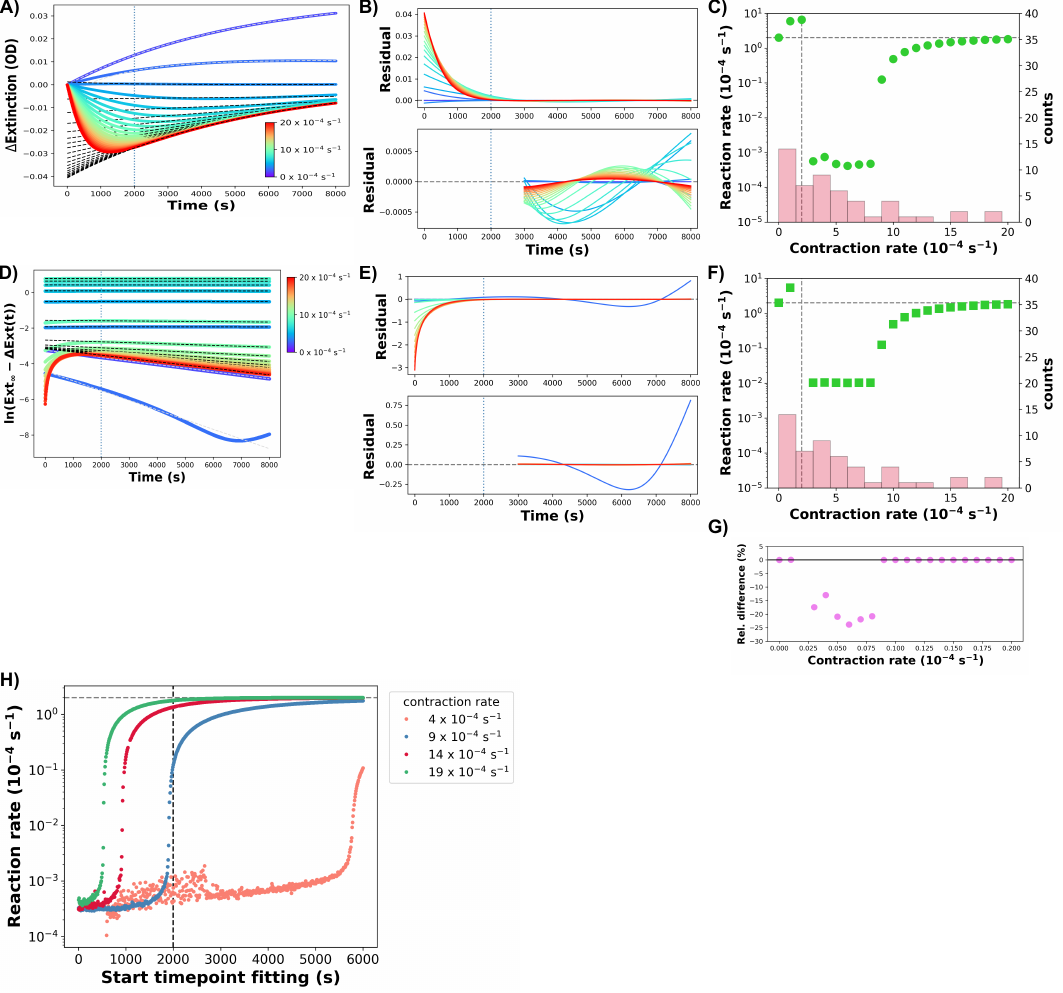}
  \caption{\textbf{Effect contraction rate on kinetics measurements in a homogeneous, contracting cavity.}
  \textbf{A)} Simulated reaction time traces for $\mathrm{k_{contraction}^{cavity}} = 0 - 20\times10^{-4} \ \text{s}^{-1}$ together with exponential fits (black dashed lines). Fitting performed between 2000 and 8000 s.
  \textbf{B)} Residuals of the exponential fits of panel A. Upper: full simulation time window; lower: zoom on the 3000-8000~s time window.
  \textbf{C)} Cavity reaction rate (green circles) extracted from direct exponential fitting vs cavity contraction rate. Experimentally observed distribution of cavity contraction rates shown in transparent red (see Figure 2C).
  \textbf{D)} Linearized reaction time traces for $\mathrm{k_{contraction}^{cavity}} = 0 - 20\times10^{-4} \ \text{s}^{-1}$ together with linear fit curves (black dashed lines). Fitting performed between 2000 and 8000~s.
  \textbf{E)} Residuals of the exponential fits of panel D. Upper: full simulation time window; lower: zoom on the 3000-8000~s time window.
  \textbf{F)} Cavity reaction rate (green squares) extracted from linearized trace fitting vs cavity contraction rate. Experimentally observed distribution of cavity contraction rates shown in transparent red (see Figure 2C).
  \textbf{G)} Relative difference between reaction rates obtained by direct trace fitting (panel C) and linearized trace fitting (panel F).
  \textbf{H)} Influence of the starting point of the fit window on extracted reaction rate for 4 selected cavity contraction rates. 
  Data of simulations with a constant cavity relative contraction = 0.5 \% and $\mathrm{k_{reaction}} = 2\times10^{-4} \ \text{s}^{-1}$. }
  \label{fig:S5}
\end{figure}

\begin{figure}[ht]
  \includegraphics[width=0.95\textwidth]{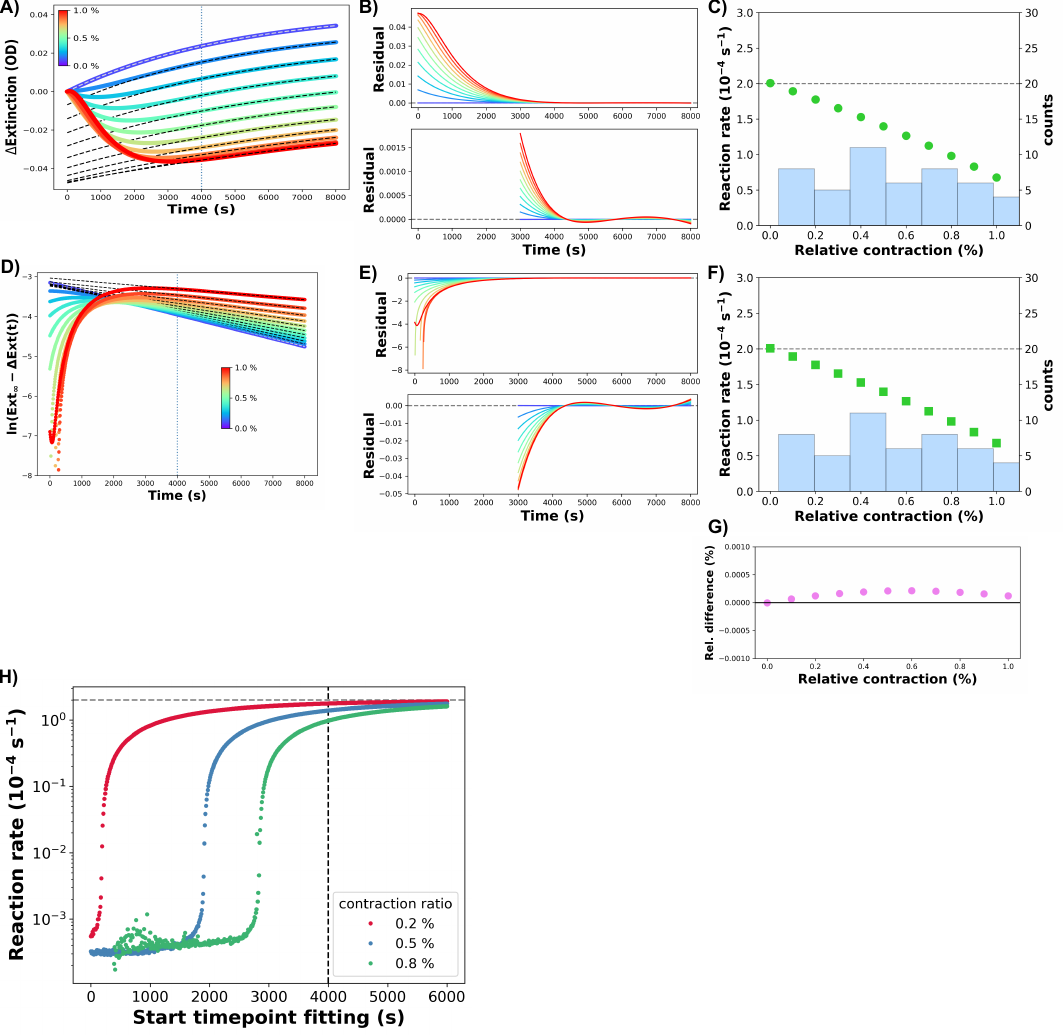}
  \caption{\textbf{Effect relative contraction on kinetics measurements in a homogeneous, contracting cavity.} 
  \textbf{A)} Simulated reaction time traces for relative contractions of 0 to 1 \% together with exponential fits (black dashed lines). Fitting performed between 4000 and 8000~s.
  \textbf{B)} Residuals of the exponential fits of panel A. Upper: full simulation time window; lower: zoom on the 3000-8000~s time window.
  \textbf{C)} Cavity reaction rate (green circles) extracted from direct exponential fitting vs cavity relative contraction. Experimentally observed distribution of cavity relative contractions shown in transparent blue (see Figure 2C).
  \textbf{D)} Linearized reaction time traces for relative contractions of 0 to 1 \% together with linear fit curves (black dashed lines). Fitting performed between 4000 and 8000~s.
  \textbf{E)} Residuals of the exponential fits of panel D. Upper: full simulation time window; lower: zoom on 3000-8000 s time window.
  \textbf{F)} Cavity reaction rate (green squares) extracted from linearized trace fitting vs cavity contraction rate. Experimentally observed distribution of cavity relative contractions shown in transparent blue (see Figure 2C).
  \textbf{G)} Relative difference between reaction rates obtained by direct trace fitting (panel C) and linearized trace fitting (panel F).
  \textbf{H)} Influence of the starting point of the fit window on extracted reaction rate for 3 selected cavity relative contractions. 
  Data of simulations with a constant $\mathrm{k_{contraction}^{cavity}} = 9\times10^{-4} \ \text{s}^{-1}$  and $\mathrm{k_{reaction}} = 2\times10^{-4} \ \text{s}^{-1}$.}
  \label{fig:S6}
\end{figure}

\begin{figure}[ht]
  \includegraphics[width=0.95\textwidth]{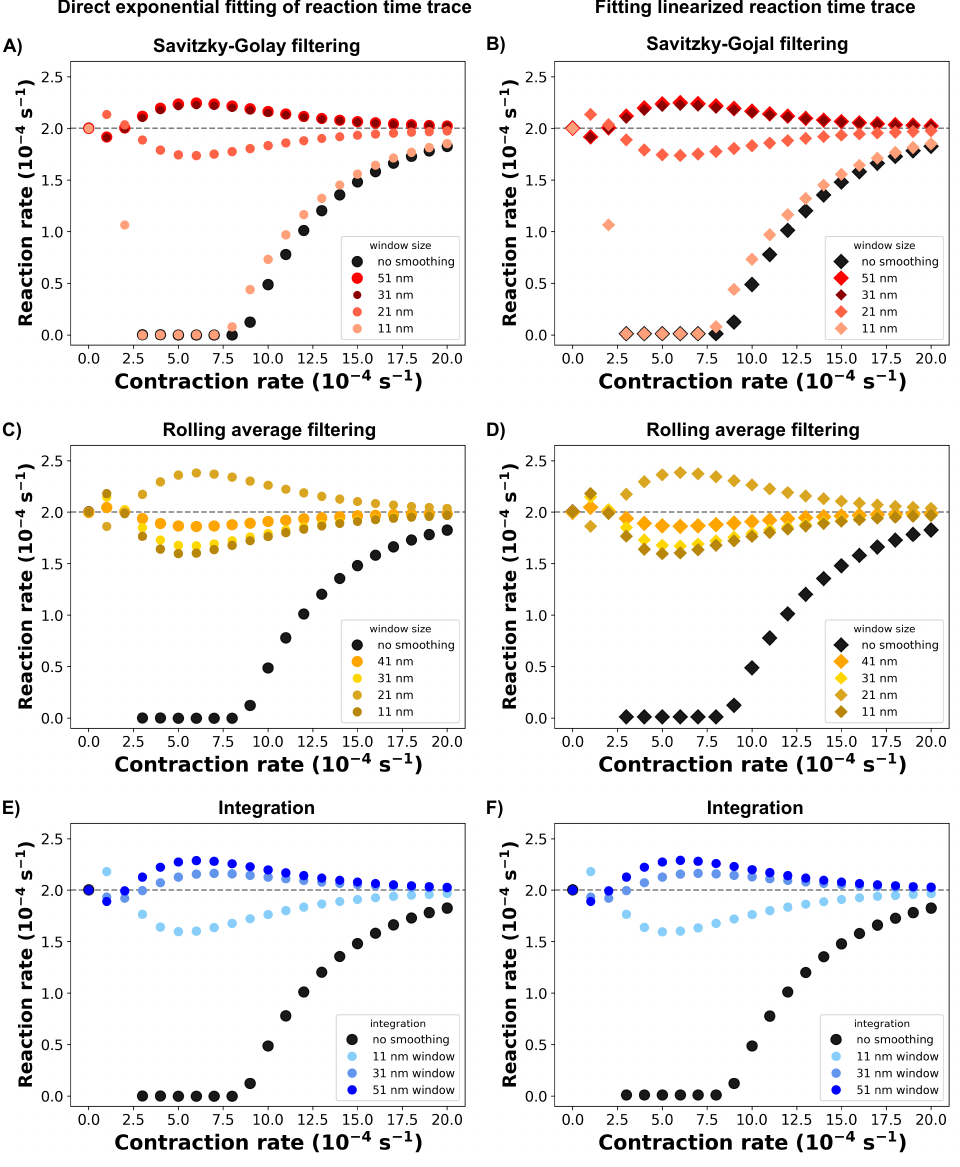}
  \caption{\textbf{Data smoothing and integration techniques applied to homogeneous cavities contracting at varying rates.}  
  Comparison of reaction rates obtained by direct exponential fitting of the reaction time trace and by fitting of the linearized reaction trace. Results of the analysis of the unsmoothed data are plotted in black. 
  \textbf{A)} and \textbf{B)} Comparison for Savitzky-Golay filtering with different filter window sizes.
  \textbf{C)} and \textbf{D)} Comparison for rolling average filtering with different filter window sizes.
  \textbf{E)} and \textbf{F)} Comparison for spectrum integration on different spectrum windows.
  Data of simulations with a constant cavity relative contraction = 0.5 \% and $\mathrm{k_{reaction}} = 2\times10^{-4} \ \text{s}^{-1}$. The fit window for both types of fitting was from 2000 to 8000 s.}
  \label{fig:S7}
\end{figure}

\begin{figure}[ht]
  \includegraphics[width=0.95\textwidth]{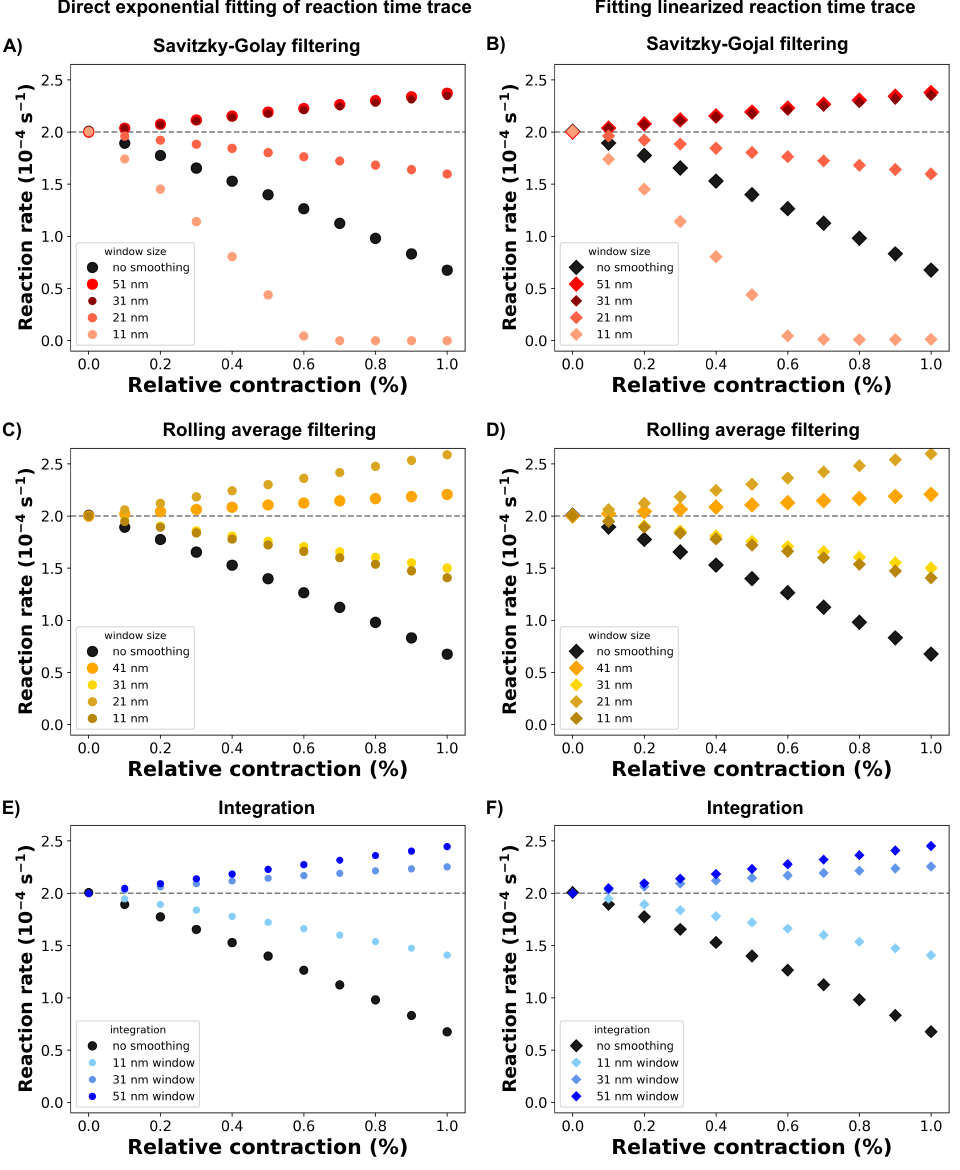}
  \caption{\textbf{Data smoothing and integration techniques applied to homogeneous cavities contracting to varying extents.} 
  Comparison of reaction rates obtained by direct exponential fitting of the reaction time trace and by fitting of the linearized reaction trace. Results of the analysis of the unsmoothed data are plotted in black. 
  \textbf{A)} and \textbf{B)} Comparison for Savitzky-Golay filtering with different filtering window sizes.
  \textbf{C)} and \textbf{D)} Comparison for rolling average filtering on different filtering window sizes.
  \textbf{E)} and \textbf{F)} Comparison for spectrum integration on different spectral windows.
  Data of simulations with a constant $\mathrm{k_{contraction}^{cavity}} = 9\times10^{-4} \ \text{s}^{-1}$  and $\mathrm{k_{reaction}} = 2\times10^{-4} \ \text{s}^{-1}$. The fit window for both types of fitting was from 2000 to 8000 s.}
  \label{fig:S8}
\end{figure}

\begin{figure}[t]
  \includegraphics[width=0.97\textwidth]{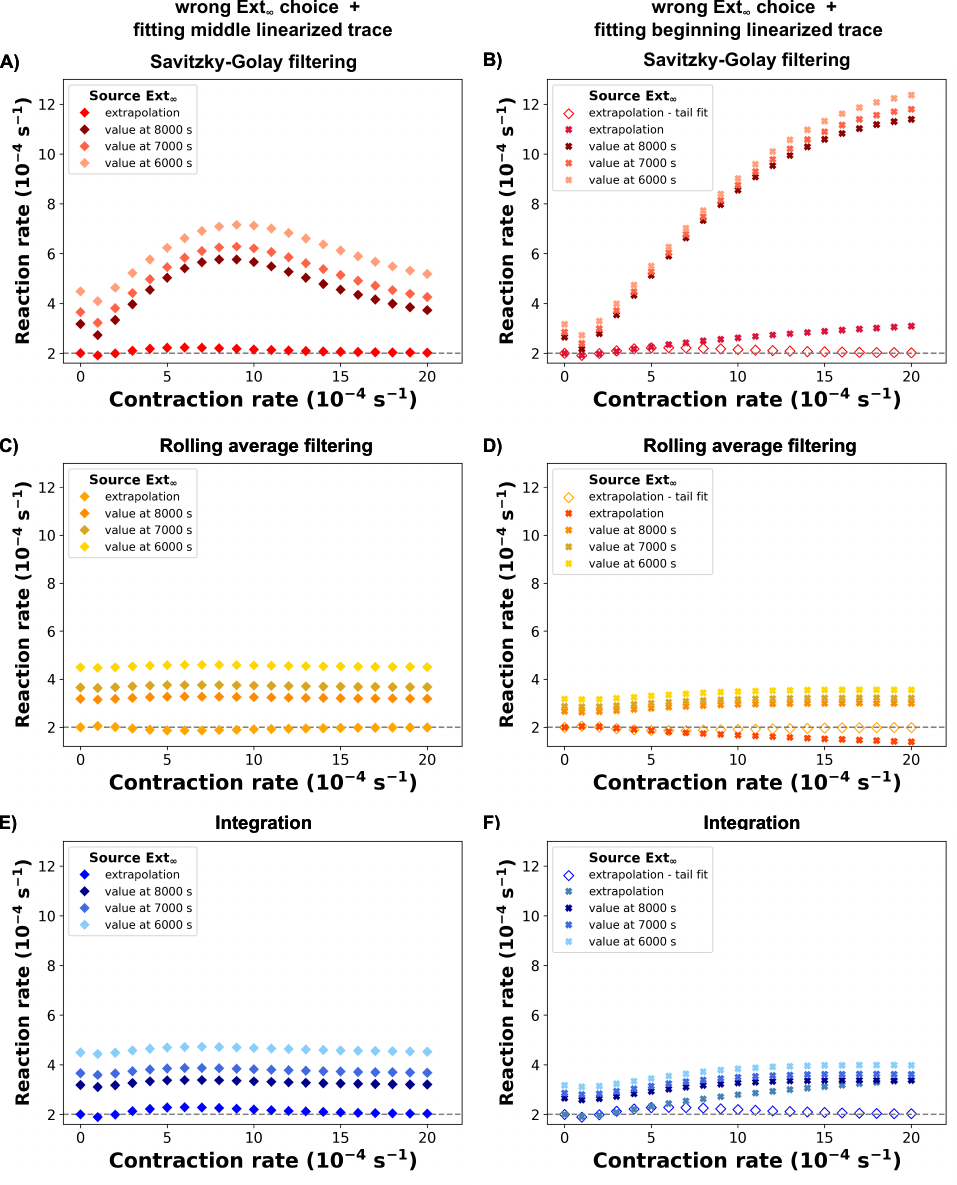}
  \caption{\textbf{Combination of data smoothing, $\mathrm{Ext_\infty}$ and fit window choice on linearized trace analysis of contracting Fabry-Perot cavities: effect of contraction rate.}
  After smoothing/integration of the simulated spectra, the linearized time traces were computed using the differential extinction values at 6000, 7000 or 8000 s, instead of using an extrapolated value (see Fig. \ref{fig:S7}). Fitting was then performed on either the beginning (0-2000~s) or the middle part (2000-4000~s) of the linearized trace. 
  \textbf{A)} and \textbf{B)} Analysis after Savitzky-Golay filtering (31 point window). 
  \textbf{C)} and \textbf{D)} Analysis after rolling average filtering (21 point window).
  \textbf{E)} and \textbf{F)} Analysis after spectrum integration (400-450 nm region).
  Data of simulations with a constant cavity relative contraction = 0.5 \% and $\mathrm{k_{reaction}} = 2\times10^{-4} \ \text{s}^{-1}$.
  }
  \label{fig:S9}
\end{figure}

\begin{figure}[ht]
    \includegraphics[width=0.93\textwidth]{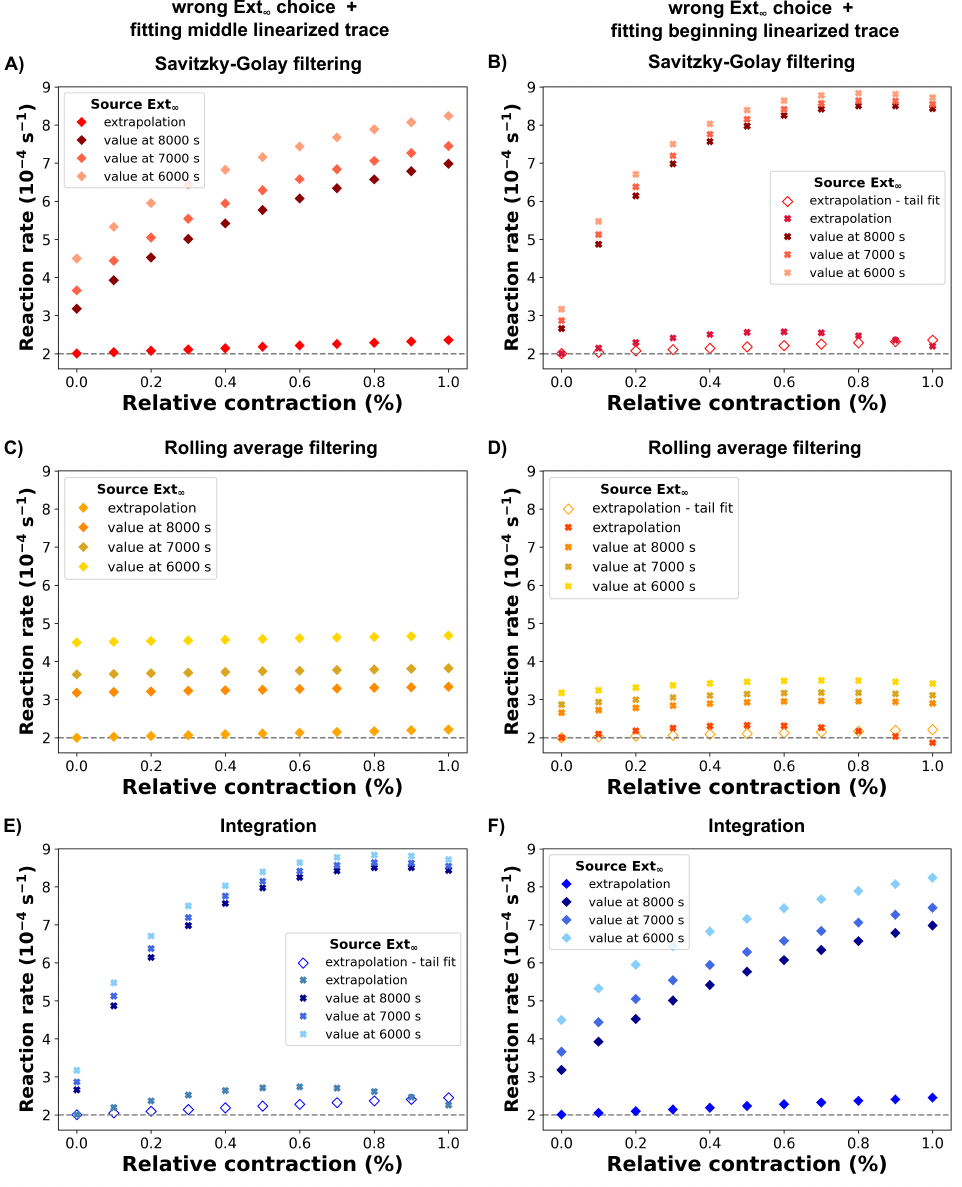}
    \caption{{\bf Combination of data smoothing, $\mathrm{Ext_\infty}$ and fit window choice on linearized trace analysis of contracting Fabry-Perot cavities: effect of relative contraction.} 
    After smoothing/integration of the simulated spectra, the linearized time traces were computed using the differential extinction values at 6000 s, 7000 s or 8000 s, instead of using an extrapolated value (see Fig. \ref{fig:S7}). Fitting was then performed on either the beginning (0-2000 s) or the middle part (2000-4000 s) of the linearized trace. 
      \textbf{A)} and \textbf{B)} Analysis after Savitzky-Golay filtering (31 point window). 
      \textbf{C)} and \textbf{D)} Analysis after rolling average filtering (21 point window).
      \textbf{E)} and \textbf{F)} Analysis after spectrum integration (400-450 nm region).
      Data of simulations with a constant $\mathrm{k_{contraction}^{cavity}} = 9\times10^{-4} \ \text{s}^{-1}$  and $\mathrm{k_{reaction}} = 2\times10^{-4} \ \text{s}^{-1}$.
    }
    \label{fig:S10}
\end{figure}

\begin{figure}[ht]
  \includegraphics[width=0.75\textwidth]{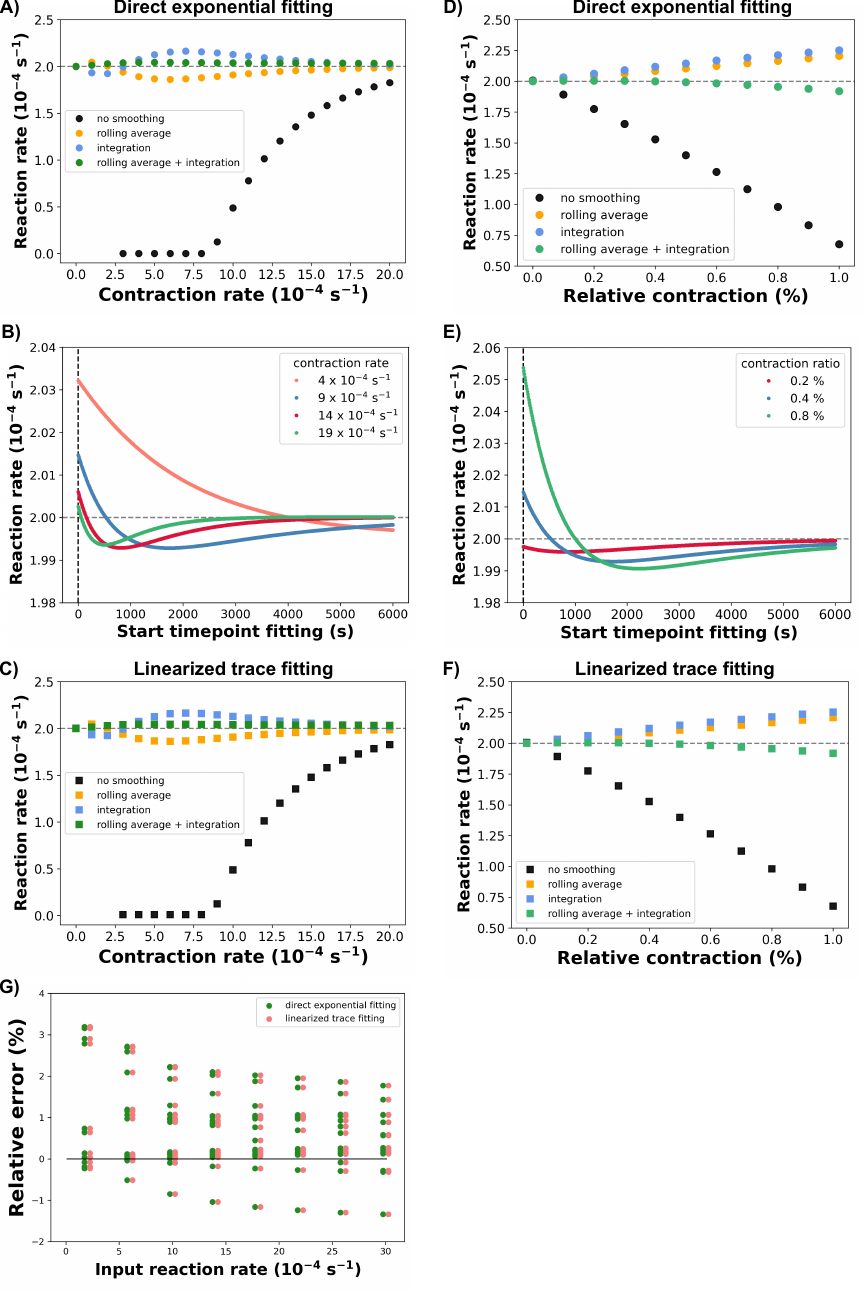}
  \caption{\textbf{Performance of integration after rolling average filtering: analysis linearized traces. }
  \textbf{A)} Reaction rates from direct trace fitting vs cavity contraction rate. Cavity relative contraction = 0.5 \%.
  \textbf{B)} Influence of starting point fit window on extracted reaction rate for 4 selected cavity contraction rates. 
  \textbf{C)} Obtained reaction rates from linearized trace fitting vs cavity contraction rate. Cavity relative contraction = 0.5 \%.
  \textbf{D)} Reaction rates from direct trace fitting vs cavity relative contraction. $\mathrm{k_{contraction}^{cavity}} = 9\times10^{-4} \ \text{s}^{-1}$
  \textbf{E)} Influence of starting point fit window on extracted reaction rate for 3 selected cavity relative contractions. 
  \textbf{F)} Reaction rates from linearized trace fitting vs cavity relative contraction. $\mathrm{k_{contraction}^{cavity}} = 9\times10^{-4} \ \text{s}^{-1}$
  Analysis by applying a rolling average filter on an 41 point window, following by integration of the 410 to 430 nm spectral region.
  }
  \label{fig:S11}
\end{figure}

\begin{figure}[t]
  \includegraphics[width=\textwidth]{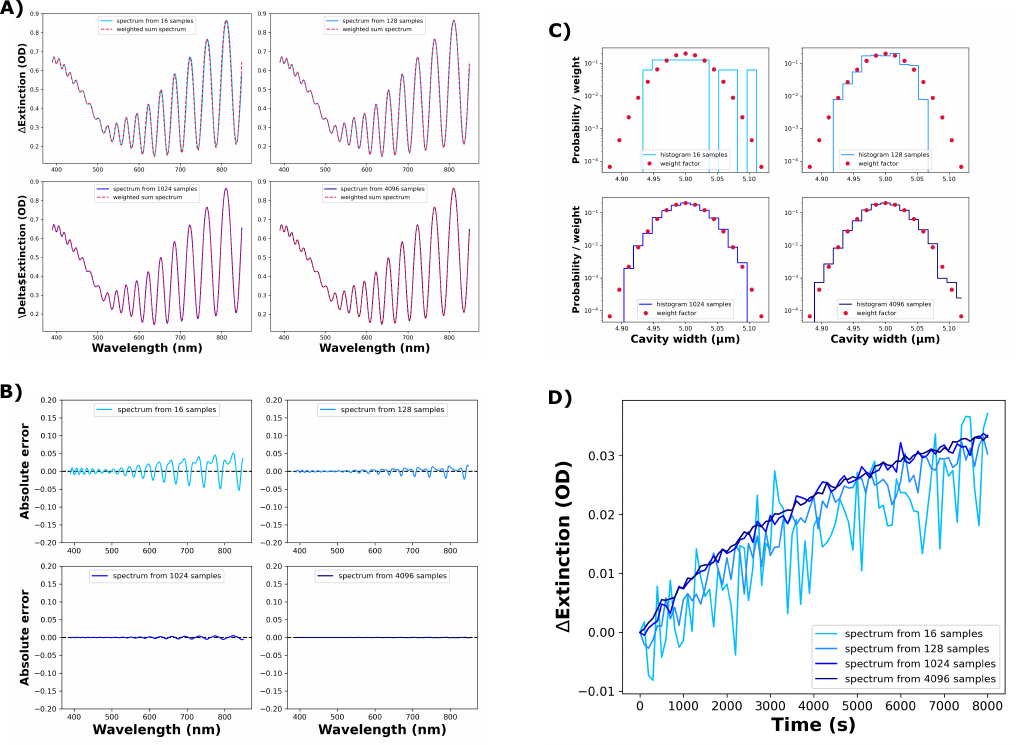}
  \caption{Simulating cavity width inhomogeneity by random sampling. 
  \textbf{A)} Simulated cavity UV/VIS spectra at t = 0 s. Cyan lines: spectra computed by averaging 16, 128, 1024, 4086 individual spectra. Red lines: reference spectrum computed using the weighted average method.\cite{Simpkins}
  \textbf{B)} Absolute differences between the spectra computed by random sampling and a reference spectrum computed by the weighted average method.
  \textbf{C)} Sampled cavity width distributions used for computing the spectra in panel A.
  \textbf{D)} Reaction time traces for an Gaussian inhomogeneous cavity simulated with random cavity width sampling. 
  Parameters: 5 \um center width and $\mathrm{L_{FWHM}}$ = 70 nm. }
  \label{fig:S12}
\end{figure}

\begin{figure}[ht]
  \includegraphics[width=\textwidth]{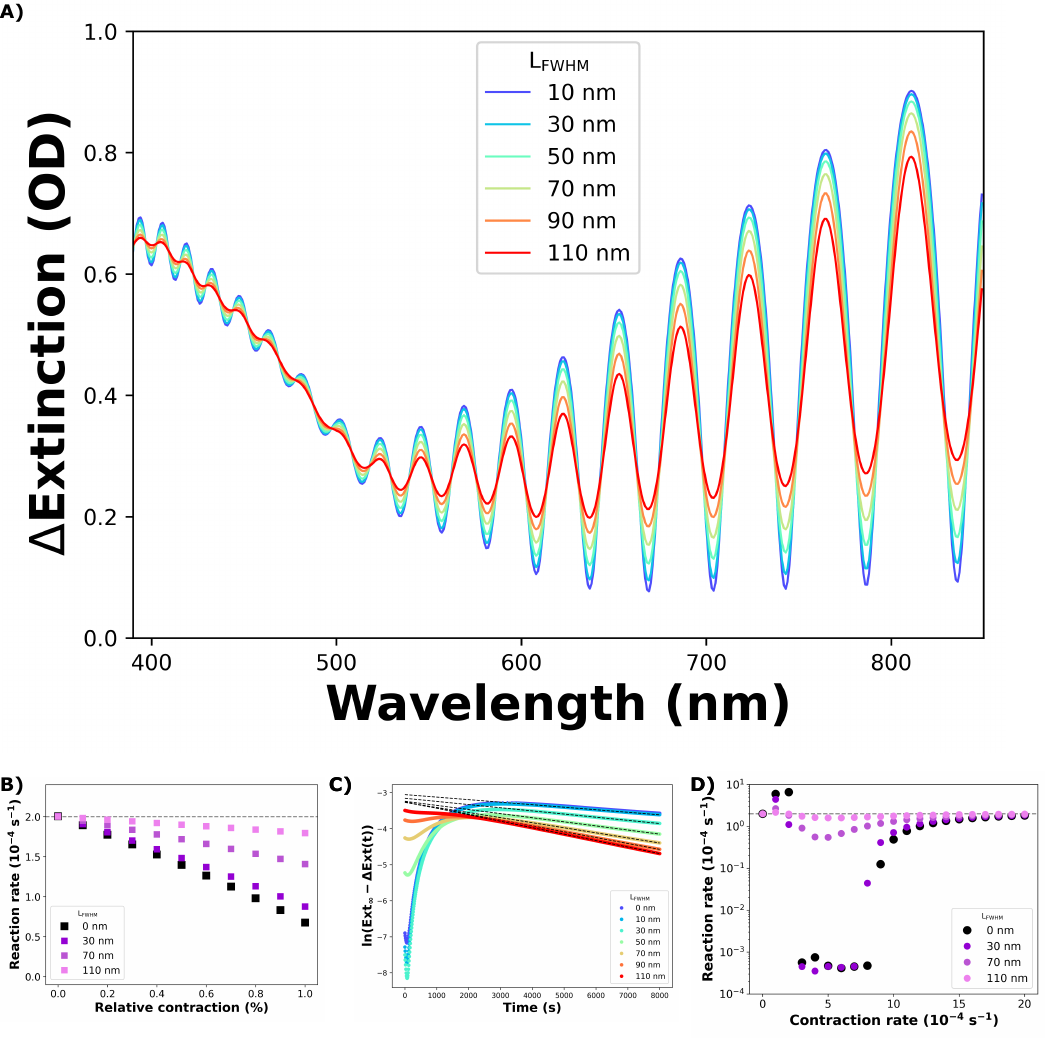}
  \caption{\textbf{Inhomogenous cavity with gaussian cavity width broadening: additional data.}
  \textbf{A)} Full spectra for a Fabry-Perot cavity with varying levels of gaussian cavity width broadening. Medium layer absorption was set to 0 corresponding to t = 0 s spectra.
  \textbf{B)} linearized form of the time traces displayed in Figure 5B together with their linear fits (dashed black lines).
  \textbf{C)} and \textbf{D)} log plot of obtained reaction rates for cavities contracting with varying contraction rates and exhibiting different levels of gaussian cavity width broadening.}
  \label{fig:S13}
\end{figure}

\begin{figure}[ht]
  \includegraphics[scale=0.85]{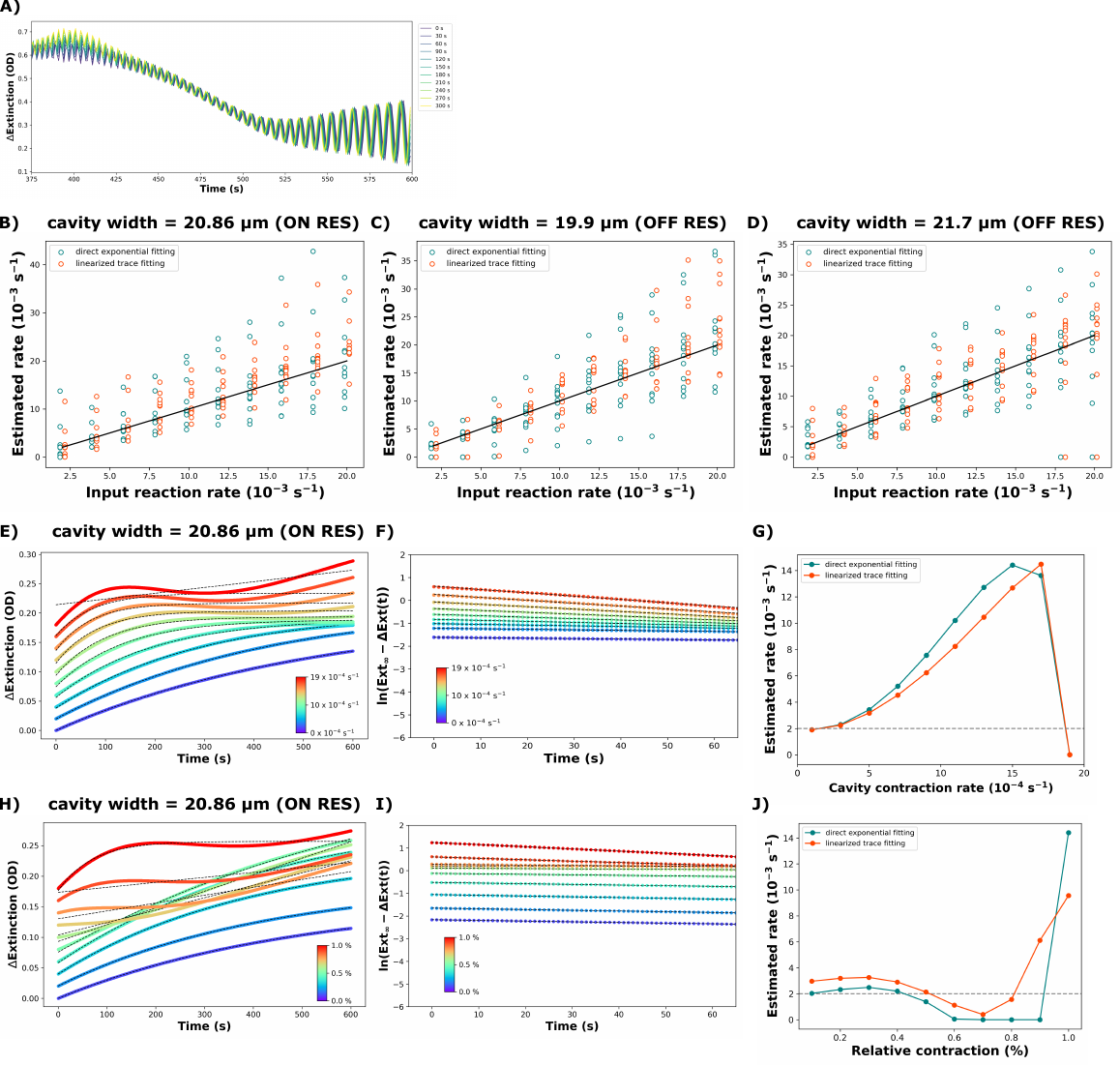}
  \caption{ Simulating the experimental data of Lather et al. (Angew Chem Intl Ed, 2019). \cite{JinoGeorge1}
  \textbf{A)} Example of simulated reaction UV/VIS spectra.
  \textbf{B-D)} Range of possible reaction rates extracted using the analysis method of Lather et al. (Angew Chem Intl Ed, 2019)~\cite{JinoGeorge1} for cavity widths of B) 20.86 \um, corresponding to an ON resonance cavity, C) 19.9 \um and D) 21.7 \um, both corresponding to OFF resonance cavities. For a given reaction rate, reaction time traces for all combinations of a set of relative contractions ($0.2, 0.6$ and $1.0 \%$) and contraction rates ($2, 8, 14$  and $20\times10^{-4} \ \text{s}{-1}$) were simulated and analyzed by fitting both the raw and linearized traces (teal and orangered curves resp.). Each dot thus represents a simulation of a cavity with one set of contraction parameters (\textit{i.e.} rel. ratio and rate).
  \textbf{E)} and \textbf{F)} Raw and linearized reaction time traces for cavities contracting 1 $\%$ at varying rates.
  \textbf{G)} Plot of the reaction rates obtained from analysis of the time traces in E) (teal) and F) (orangered).
  \textbf{H)} and \textbf{I)} Raw and linearized reaction time traces for cavities contracting to varying degrees at a rate of $15\times10^{-4} \ \text{s}^{-1}$.
  \textbf{J)} Plot of the reaction rates obtained from analysis of the time traces in H) (teal) and I) (orangered).
  Simulations in E-J) with a input reaction rate $\mathrm{k_{reaction} = 2\times10^{-3} \ \text{s}^{-1}}$.
  }
  \label{fig:S14}
\end{figure}

\begin{figure}[t]
  \includegraphics[scale=0.85]{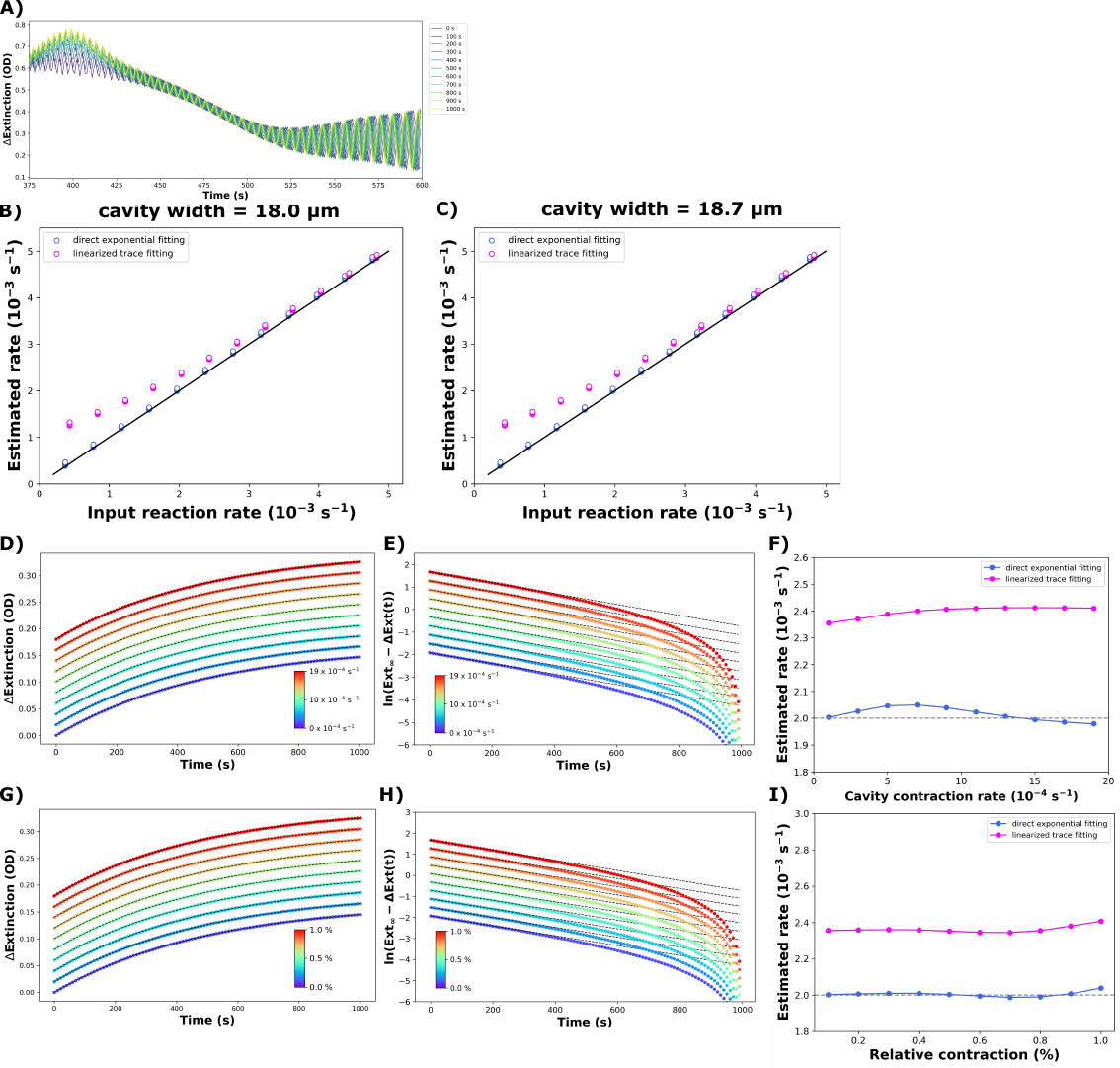}
  \caption{Simulating the experimental data of Singh et al. (ChemPhysChem, 2023). \cite{JinoGeorge2}
  \textbf{A)} Example of simulated reaction UV/VIS spectra.
  \textbf{B)} \& \textbf{C)} Range of possible reaction rates extracted using the analysis method of Singh et al. (ChemPhysChem, 2023)~\cite{JinoGeorge2} for cavity widths of B) 18.0 \um and C) 18.7 \um. 
  For each reaction rate, reaction time traces for all combinations of a set of relative contractions ($0.2, 0.6$ and $1.0 \%$) and contraction rates ($2, 8, 14$  and $20\times10^{-4} \ \text{s}{-1}$) were simulated and analyzed by fitting both the raw and linearized traces (blue and pink curves resp.). Each dot thus represents a simulation of a cavity with one set of contraction parameters (\textit{i.e.} rel. ratio and rate).
  \textbf{D)} and \textbf{E)} Raw and linearized reaction time traces for cavities contracting 1 $\%$ at varying rates.
  \textbf{F)} Plot of the reaction rates obtained from analysis of the time traces in C) (blue) and D) (pink).
  \textbf{G)} and \textbf{H)} Raw and linearized reaction time traces for cavities contracting to varying degrees at a rate of $15\times10^{-4} \ \text{s}^{-1}$.
  \textbf{I)} Plot of the reaction rates obtained from analysis of the time traces in F) (blue) and G) (pink).
  Simulations in D-I) with a input reaction rate $\mathrm{k_{reaction} = 2\times10^{-3} \ \text{s}^{-1}}$.
  }
  \label{fig:S15}
\end{figure}

\begin{figure}[t]
  \includegraphics[scale=0.85]{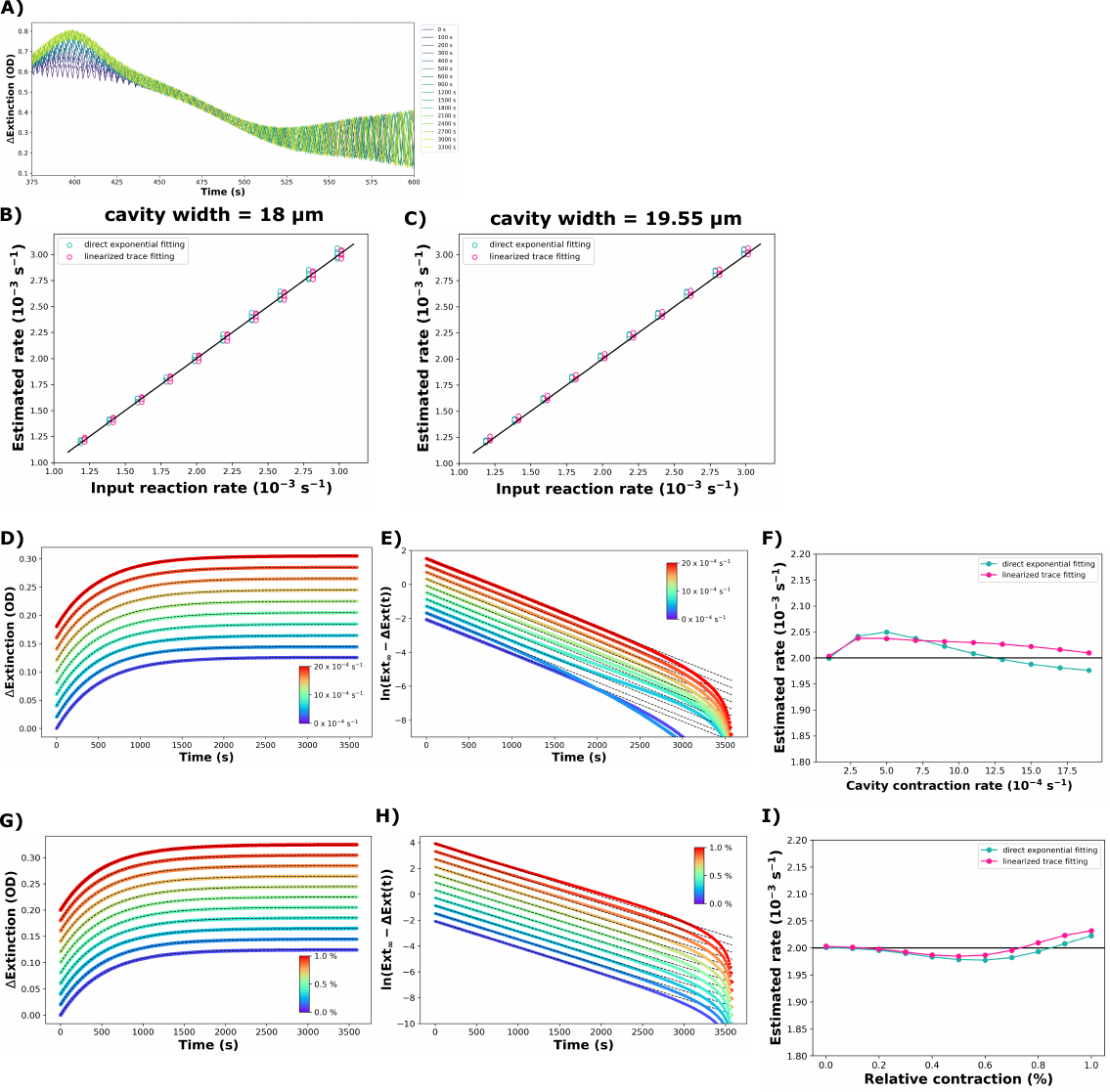}
  \caption{Simulating the experimental data of Weisehan and Xiong (J Chem Phys, 2021).\cite{WeiXiong}
  \textbf{A)} Example of simulated reaction UV/VIS spectra.
  \textbf{B)} \& \textbf{C)} Range of possible reaction rates extracted using the analysis method of Weisehan and Xiong (J Chem Phys, 2021)~\cite{WeiXiong} for cavity widths of B) 18.0 \um and C) 19.55 \um. 
  For each reaction rate, reaction time traces for all combinations of a set of relative contractions ($0.2, 0.6$ and $1.0 \%$) and contraction rates ($2, 8, 14$  and $20\times10^{-4} \ \text{s}{-1}$) were simulated and analyzed by fitting both the raw and linearized traces (green and pink curves resp.). Each dot thus represents a simulation of a cavity with one set of contraction parameters (\textit{i.e.} rel. ratio and rate).
  \textbf{D)} and \textbf{E)} Raw and linearized reaction time traces for cavities contracting 1 $\%$ at varying rates.
  \textbf{F)} Plot of the reaction rates obtained from analysis of the time traces in C) (green) and D) (pink).
  \textbf{G)} and \textbf{H)} Raw and linearized reaction time traces for cavities contracting to varying degrees at a rate of $15\times10^{-4} \ \text{s}^{-1}$.
  \textbf{I)} Plot of the reaction rates obtained from analysis of the time traces in F) (green) and G) (pink).
  Simulations in D-I) with a input reaction rate $\mathrm{k_{reaction} = 2\times10^{-3} \ \text{s}^{-1}}$.
  }
  \label{fig:S16}
\end{figure}

\clearpage

\bibliography{bib_VSCanalysis}